\documentclass[a4paper,fleqn,usenatbib]{mnras}
\usepackage{newtxtext,newtxmath}
\usepackage[T1]{fontenc}
\usepackage{ae,aecompl}

\usepackage{graphicx}	
\usepackage{amsmath}	
\usepackage{amssymb}	

\newcommand{\teff}{\tau_\mathrm{eff}}

\defcitealias{Kakiichi18}{K18}

\title[UVB bias and ghost proximity effect]{Ionization bias and the ghost proximity effect near $z\gtrsim6$ quasars in the shadow of proximate absorption systems}

\author[F. B. Davies]{ 
Frederick B. Davies,$^{1}$\thanks{E-mail: davies@physics.ucsb.edu}
\\
$^{1}$Department of Physics, University of California, Santa Barbara, CA 93106-9530, USA
}

\date{Accepted XXX. Received YYY; in original form ZZZ}

\pubyear{2019}

\begin{document}
\label{firstpage}
\pagerange{\pageref{firstpage}--\pageref{lastpage}}
\maketitle

\begin{abstract}
The larger-than-expected scatter in the opacity of the Ly$\alpha$ forest suggests that the metagalactic ionizing background is strongly fluctuating at $z > 5.5$. Models for ionizing background fluctuations predict a strong positive bias on large scales, so the environments of massive $>10^{12}\,{\rm M}_\odot$ dark matter halos, e.g. $z\sim6$ quasar hosts, would be ideal laboratories to constrain the sources of ionizing photons. While the quasars themselves should overwhelm any plausible ionizing photon contribution from neighboring galaxies, proximate damped Ly$\alpha$ absorbers (DLAs) have recently been discovered in the foreground of $z\sim6$ quasars, and the Ly$\alpha$ forest in the shadow of these DLAs could probe the local ionization environment. Using Gpc$^3$ simulations of $z=6$ ionizing background fluctuations, we show that while the Ly$\alpha$ forest signal from ionization bias around a quasar host halo should be strong, it is likely suppressed by the associated intergalactic matter overdensity. We also show that the quasar itself may still overwhelm the clustering signal via a ``ghost" of the proximity effect from the quasar radiation causing a large-scale bias in the ionizing photon mean free path. This ghost proximity effect is sensitive to the lifetime and geometry of quasar emission, potentially unlocking a new avenue for constraining these fundamental quasar properties. Finally, we present observations of a $z\sim6$ quasar with a proximate DLA which shows a strong excess in Ly$\alpha$ forest transmission at the predicted location of the ghost proximity effect.
\end{abstract}


\section{Introduction}

The nature of the sources which reionized the Universe, and continued to keep it ionized afterwards, is at the forefront of current high-redshift cosmological investigation. While luminous quasars are obvious candidates due to their strong ionizing spectrum which is regularly observed directly at lower redshift (e.g. \citealt{Prochaska09,Worseck14,Stevans14,Lusso15}), they become too scarce at $z\gtrsim5$ to reionize the Universe (e.g. \citealt{Jiang16,Matsuoka18}) or maintain the highly ionized state of the intergalactic medium (IGM) after reionization is complete (e.g. \citealt{HM12,Khaire16,McGreer18}). Thus young stars in star-forming galaxies are the most widely accepted source of ionizing photons, but direct detection of ionizing photons escaping from galaxies is extremely challenging and limited to $z<4$ due to IGM absorption (e.g. \citealt{Mostardi15,Vanzella16,Steidel18}), so the actual ionizing photon budget at $z\gtrsim4$ is highly uncertain.

The metagalactic ionizing radiation field, known as the ionizing or UV background, is believed to be strongly fluctuating at $z\gtrsim5.5$ due to the greater-than-expected fluctuations in the opacity of the Ly$\alpha$ forest \citep{Becker15,Chardin15,Chardin17,DF16,D'Aloisio18,Becker18}. These fluctuations could arise due to a relatively short mean free path of ionizing photons ($\lambda_{\rm mfp}$) and the strong clustering of the sources of ionizing photons at these redshifts \citep{MF09}, and they could be enhanced by residual neutral patches from incomplete reionization \citep{Kulkarni19}. \citet[][henceforth \citetalias{Kakiichi18}]{Kakiichi18} quantified the bias of ionizing radiation around Lyman-break galaxies using halo model arguments and an approximate one-dimensional radiative transfer scheme, and showed that the clustering of ionizing photon sources leads to a strong positive bias in Ly$\alpha$ forest transmission at $z\sim6$, i.e. there should be excess Ly$\alpha$ forest transmission in the spectra of background quasars within several proper Mpc of luminous galaxies due to the associated overdensity of nearby faint sources (see also \citealt{Davies17b}). Measurements of this ionizing background bias may then have the potential to identify the nature of the sources of ionizing photons at the end of the reionization epoch and constrain the escape fraction of ionizing photons.

\begin{figure*}
\begin{center}
\resizebox{16cm}{!}{\includegraphics[trim={1em 4em 1em 5em},clip]{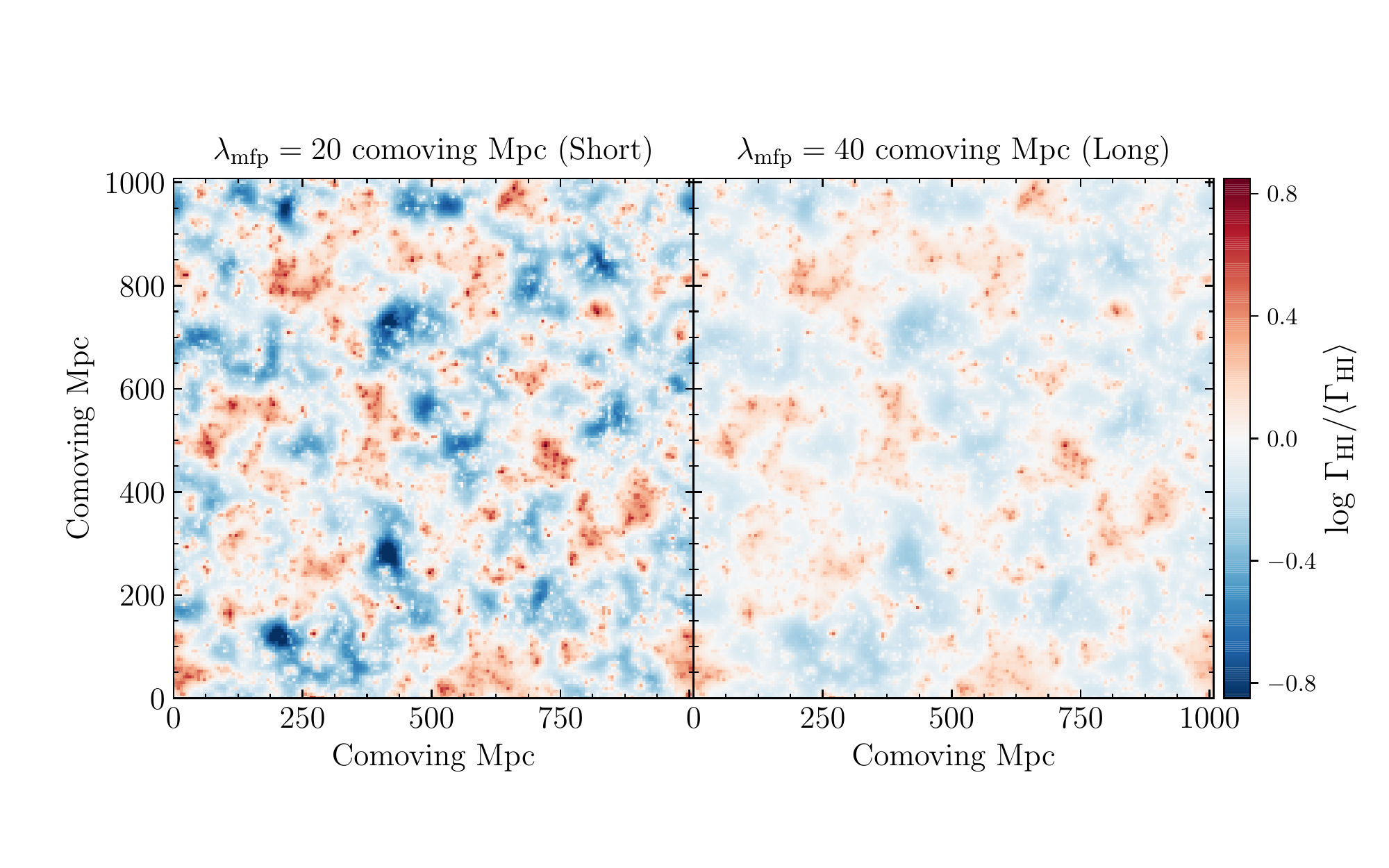}}
\end{center}
\caption{Slices through the ionizing background models with $\lambda_{\rm mfp}=20$ (left; ``short") and 40 (right; ``long") comoving Mpc.}
\label{fig:uvb_slices}
\end{figure*}

The ionizing background bias signal should be the strongest around the most massive, and thus most biased, dark matter halos. While \citetalias{Kakiichi18} employed Lyman-break galaxies for this purpose, the host halos of $z\sim6$ quasars are potentially even more massive, and thus would be ideal locations to search for enhanced Ly$\alpha$ forest transmission due to a biased ionizing background. Unfortunately, the quasars themselves produce far more ionizing photons than the surrounding galaxies, completely overwhelming the clustering signal. However, if the ionizing radiation emitted by the quasar is blocked by an absorption system which is optically thick at the Lyman limit, then in principle the Ly$\alpha$ forest beyond the saturated trough of the absorber could allow for measurements of the local bias in the ionizing background. 

In this work, we construct a semi-numerical model for ionizing background fluctuations around the most massive dark matter halos at $z\sim6$, building upon the models from \citet{DF16} and \citet{Davies17b}. In \S~\ref{sec:bias} we describe the construction of the model, and present the resulting ionizing background fluctuations and corresponding Ly$\alpha$ forest transmission signal around halos. In \S~\ref{sec:ghost} we show that the \emph{ghost proximity effect}, a direct consequence of mean free path fluctuations, is likely to be much stronger than the halo bias signal, and has the potential to constrain the duration and anisotropy of quasar emission. In \S~\ref{sec:j056} we show preliminary evidence for the ghost proximity effect in a newly identified proximate DLA quasar at $z\sim6$. Finally, we conclude in \S~\ref{sec:conc}.

We assume a $\Lambda$CDM cosmology with $h=0.7$, $\Omega_m=0.3$, $\Omega_\Lambda=0.7$, $\Omega_b=0.047$, $\sigma_8 = 0.8$, and $n_s=0.96$.

\section{The biased ionizing background around massive halos} \label{sec:bias}

Due to their extreme rarity ($\lesssim1\,{\rm Gpc}^{-3}$, e.g. \citealt{Jiang16}), the dark matter halos hosting luminous quasars at $z\ga6$ are thought to be very massive, although they may not be the \emph{most} massive halos \citep{Fanidakis13}, particularly if they have short duty cycles \citep{CG18}. Simulating the ionizing background fluctuations around these most massive halos, and the variations from halo to halo, thus requires an extremely large Gpc-scale volume. In this section, we briefly summarize our semi-numerical simulations of the ionizing background, which closely follow the method described in \citet{DF16}, and the basic properties of the resulting fluctuations around massive dark matter halos.

\subsection{Semi-numerical ionizing background fluctuations in a Gpc$^3$ volume} \label{sec:uvb}

We first instantiated a set of $5760^3$ cosmological initial conditions in a volume 1008 comoving Mpc on a side using the semi-numerical reionization code \texttt{21cmFAST} \citep{Mesinger11}, and used the excursion set halo finder in \citep{MF07} to populate the volume with dark matter halos down to a minimum mass of $2\times10^9\,{\rm M}_\odot$. A $1440^3$ evolved density field at $z=6$ was then computed using the Zel'dovich approximation \citep{Zel'dovich70}, the excursion set halos were shifted using the same displacement field \citep{MF07}. Galaxy UV luminosities were assigned to the dark matter halos via abundance matching (e.g. \citealt{VO04}) the dark matter halos to the UV luminosity function from \citet{Bouwens15}. Ionizing luminosities were then assigned to galaxies assuming a fixed (i.e. not dependent on halo mass) conversion factor from their (non-ionizing) UV luminosity.

The ionizing radiation field was computed as in \citet{DF16} on a $180^3$ grid for the fiducial simulations and $144^3$ for an expanded suite of simulations described later in \S~\ref{sec:ghost}. The sources of ionizing photons, i.e. the galaxies mentioned above, were smoothed onto the same grid to create an ionizing emissivity field. The ionizing opacity of each cell, as parametrized by the mean free path of ionizing photons $\lambda_{\rm mfp}$, was assumed to follow the relationship\footnote{The density dependence assumed here may be too strong; \citet{Chardin17} found $\lambda \propto \Delta^{-0.4}$ via post-processing the Sherwood hydrodynamical simulations \citep{Bolton17}. Adopting this shallower dependence would modestly increase the strength of the ionizing background fluctuations at fixed $\lambda_{\rm mfp}$.} $\lambda \propto \Gamma_{\rm HI}^{2/3} \Delta^{-1}$, where $\Gamma_{\rm HI}$ is the photoionization rate of hydrogen and $\Delta$ is the matter overdensity. In the following, the mean free path at the average $\Gamma_{\rm HI}$ and mean IGM density will be denoted ``$\lambda_{\rm mfp}$," but note that the actual mean free path in the simulations is spatially variable. For computational efficiency we considered only a single energy of ionizing photons, $E_{\rm ion}\approx18$ eV, having tested that this characteristic energy closely reproduces the background fluctuations resulting from a multi-frequency treatment (assuming a power law spectrum of galaxies, $L_\nu \propto \nu^{-2}$, e.g. \citealt{BB13}). The ionizing radiation field calculation was iterated several times, typically between 7 to 10 times depending on the choice of $\lambda_{\rm mfp}$, until the radiation intensity and opacity fields were converged to $<0.1\%$. The brute force method for calculating the ionizing radiation field that we employ following \citet{DF16} is not particularly computationally efficient, however for this large, low-resolution volume the computation can be sped up considerably by only allowing radiation to propagate a maximum distance from each cell. For the fiducial calculations, each grid cell ``sees" a cubical volume with a side length of $14\times\lambda_{\rm mfp}$.

Ionizing background models were computed with $\lambda_{\rm mfp}=20$ and $40$ comoving Mpc, where the former corresponds to a short mean free path model that could explain the large Ly$\alpha$ forest fluctuations (\citealt{DF16}, although see \citealt{Kulkarni19}), while the latter more closely corresponds to (but is still somewhat less than) the extrapolation of the \citet{Worseck14} measurements of $\lambda_{\rm mfp}$ to $z=6$. We will refer to these models as ``short" and ``long" $\lambda_{\rm mfp}$ in the rest of this work. The ionizing emission from galaxies was normalized such that both models result in a mean photoionization rate of $\langle\Gamma_{\rm HI}\rangle = 2.0\times10^{-13}$ s$^{-1}$, consistent with measurements at $z\sim6$ \citep{Calverley11,WB11,Davies17}. In Figure~\ref{fig:uvb_slices} we show slices of the resulting ionizing background fluctuations for the two $\lambda_{\rm mfp}$ values. The dominant effect of changing the mean free path is a change in the amplitude of background fluctuations -- the scales of the visually apparent features in Figure~\ref{fig:uvb_slices} are almost identical between the two panels, likely fixed by the clustering scale of the ionizing sources.

\begin{figure}
\begin{center}
\resizebox{8.5cm}{!}{\includegraphics[trim={1em 1em 1em 1em},clip]{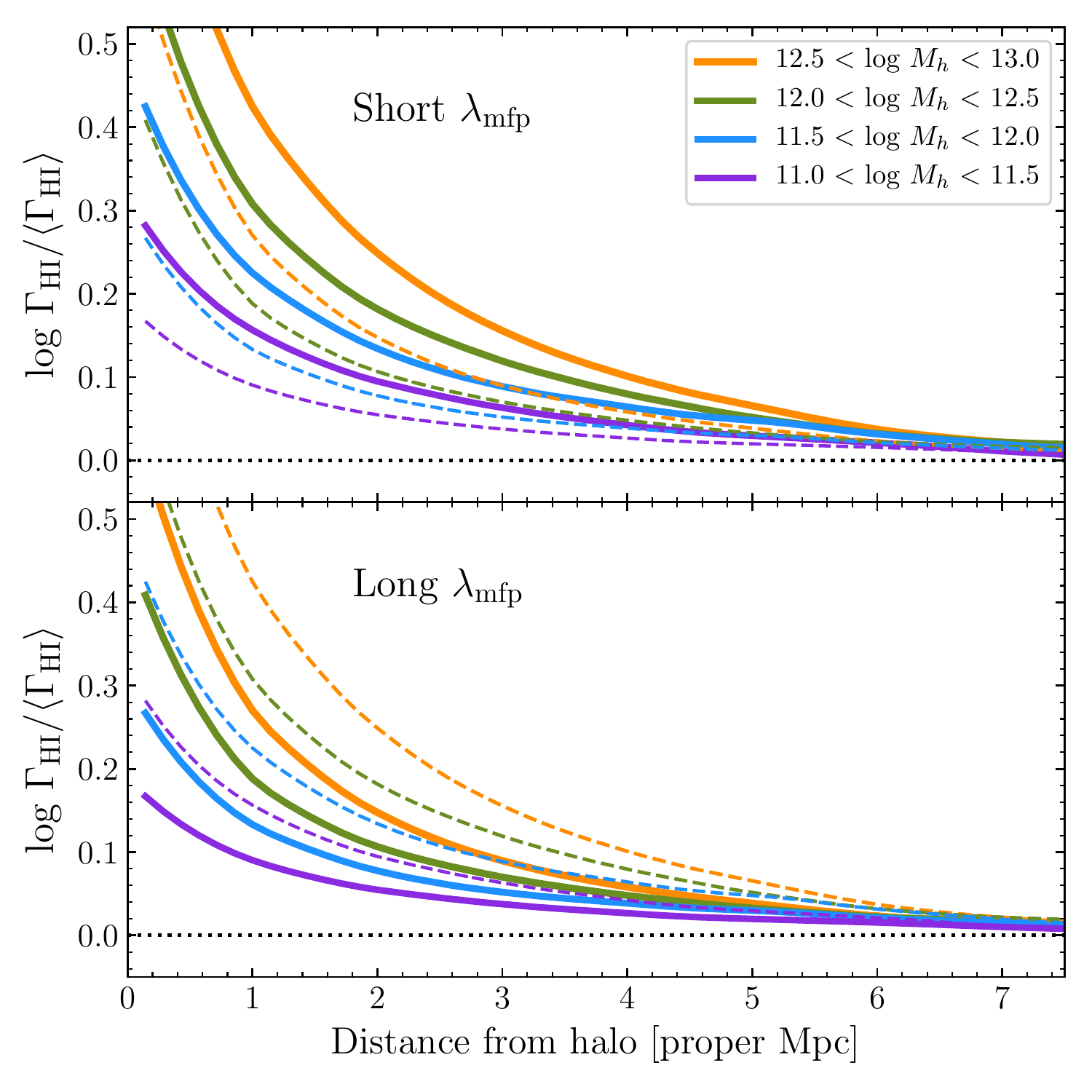}}
\end{center}
\caption{Average ionization rate along lines of sight centered on the most massive halos in the short $\lambda_{\rm mfp}$ (top) and long $\lambda_{\rm mfp}$ (bottom) models. The different colors represent different ranges of halo mass, and the dashed curves show the corresponding profiles from the other model for comparison.}
\label{fig:halo_bias}
\end{figure}

\begin{figure}
\begin{center}
\resizebox{8.5cm}{!}{\includegraphics[trim={1em 1em 1em 1em},clip]{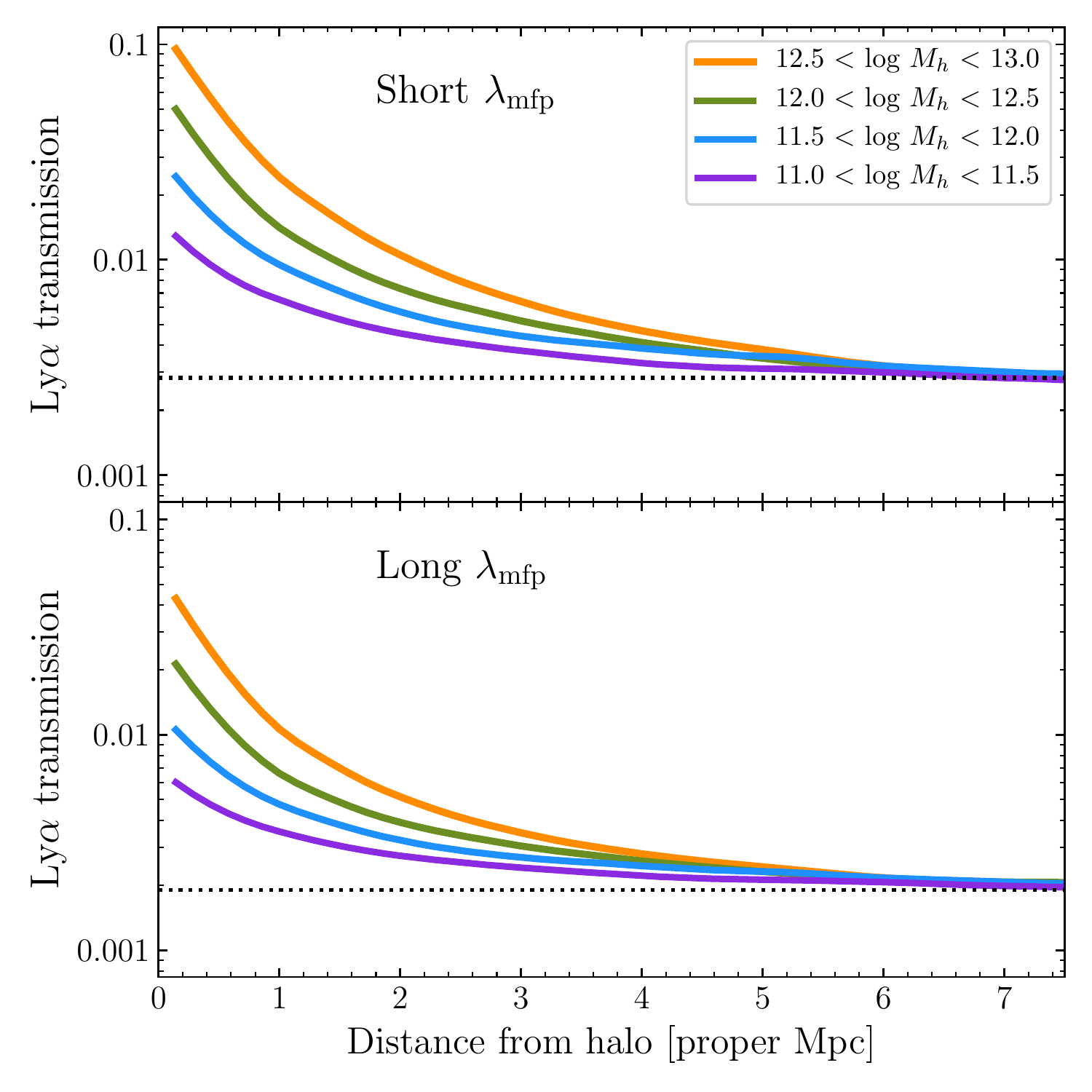}}
\end{center}
\caption{Similar to Figure~\ref{fig:halo_bias}, but now showing the average Ly$\alpha$ forest transmission profiles for massive halos in the short $\lambda_{\rm mfp}$ (top) and long $\lambda_{\rm mfp}$ (bottom) models computed via the $\tau_{\rm eff}-\Gamma_{\rm HI}$ relationship from \citet{Davies19b}. Horizontal dotted lines show the mean Ly$\alpha$ forest transmission for each model.}
\label{fig:tra_bias}
\end{figure}

\subsection{Ionization rate and Ly$\alpha$ forest transmission profiles}

To determine the biased radiation field around halos, the photoionization rate was extracted along random lines of sight centered on locations of halos in the semi-numerical simulation. At each position along the skewer, the ionization rate was interpolated via trilinear interpolation of the neighboring log $\Gamma_{\rm HI}$ values. In Figure~\ref{fig:halo_bias} we show the median $\Gamma_{\rm HI}$ profiles for different ranges of halo mass in the short $\lambda_{\rm mfp}$ (top) and long $\lambda_{\rm mfp}$ (bottom) simulations. As expected, $\Gamma_{\rm HI}$ is enhanced close to massive halos, and more massive halos are associated with a stronger $\Gamma_{\rm HI}$ enhancement. Because the shorter mean free path model exhibits stronger ionizing background fluctuations in general (cf. Figure~\ref{fig:uvb_slices}), the bias of $\Gamma_{\rm HI}$ around massive halos is stronger as well.

\begin{figure}
\begin{center}
\resizebox{8.5cm}{!}{\includegraphics[trim={1em 1em 1em 1em},clip]{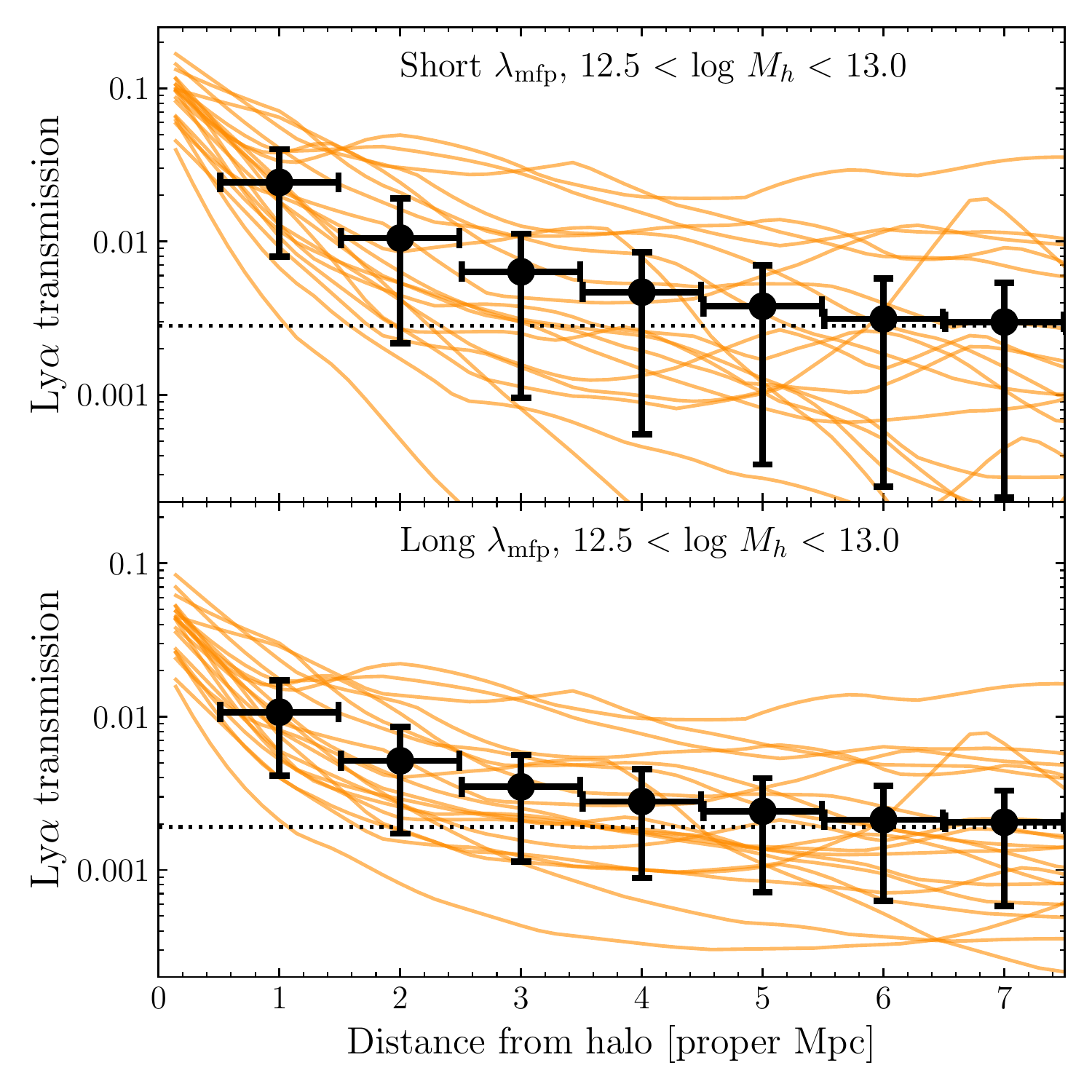}}
\end{center}
\caption{Ly$\alpha$ forest transmission profiles of individual lines of sight from massive halos ($12.5 < \log{M_h} < 13.0$) in the short $\lambda_{\rm mfp}$ (top) and long $\lambda_{\rm mfp}$ (bottom) models. Black points with error bars show the median and 16--84th percentiles of Ly$\alpha$ forest transmission from a much larger set of random skewers. Horizontal dotted lines show the mean Ly$\alpha$ forest transmission for each model.}
\label{fig:tra_bias_var}
\end{figure}

\begin{figure}
\begin{center}
\resizebox{8.5cm}{!}{\includegraphics[trim={1em 1em 1em 1em},clip]{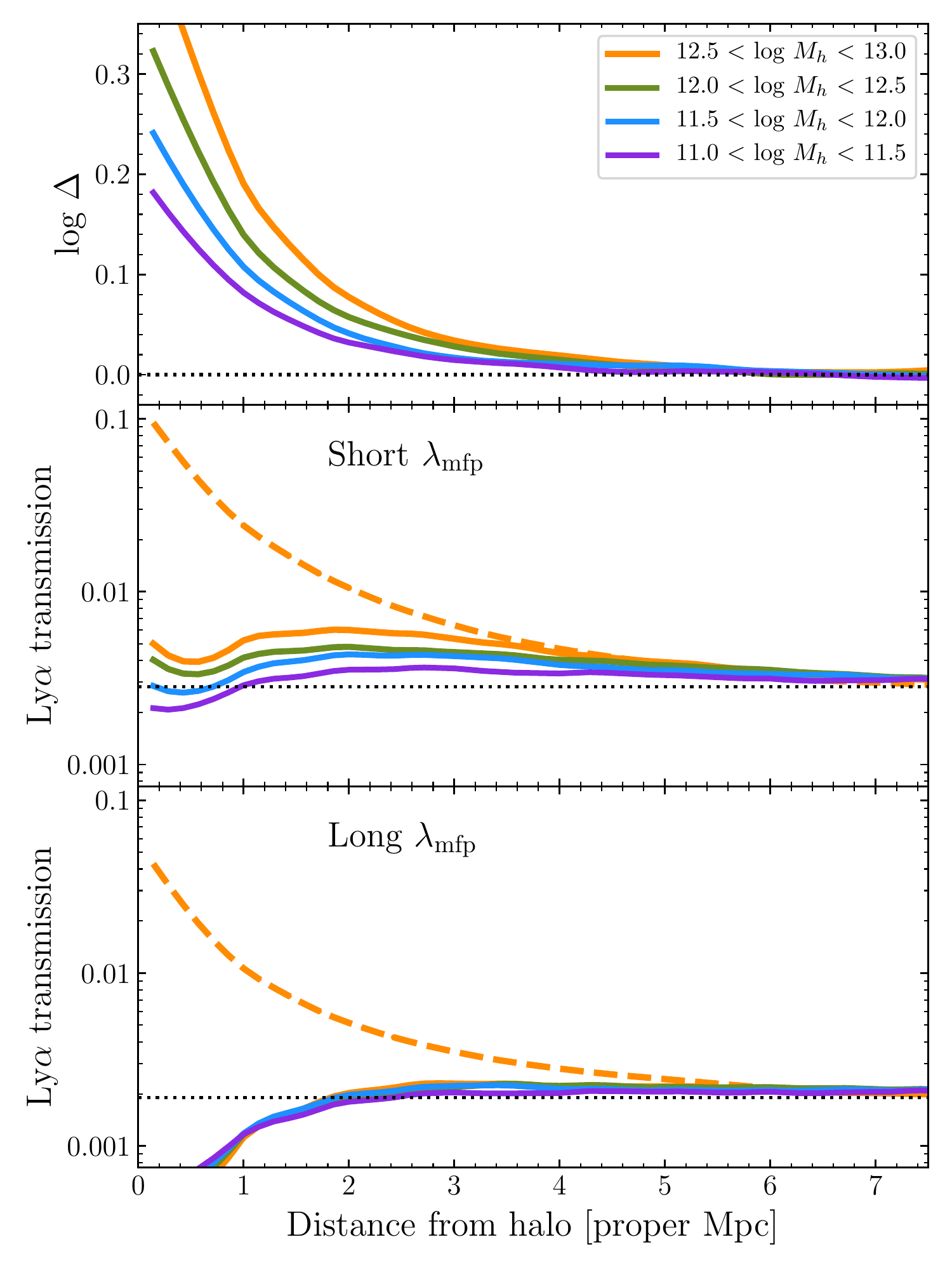}}
\end{center}
\caption{Similar to Figure~\ref{fig:tra_bias} but now including the effect of enhanced IGM density close to massive halos (top panel) on the Ly$\alpha$ forest transmission (bottom two panels) by rescaling $\teff$ proportional to $\Delta^{0.9}$. The dashed orange curves show the unmodified Ly$\alpha$ forest transmission for $12.5 < \log{M_{\rm h}} < 13.0$ halos.}
\label{fig:tra_bias2}
\end{figure}

\begin{figure}
\begin{center}
\resizebox{8.5cm}{!}{\includegraphics[trim={1em 1em 1em 1em},clip]{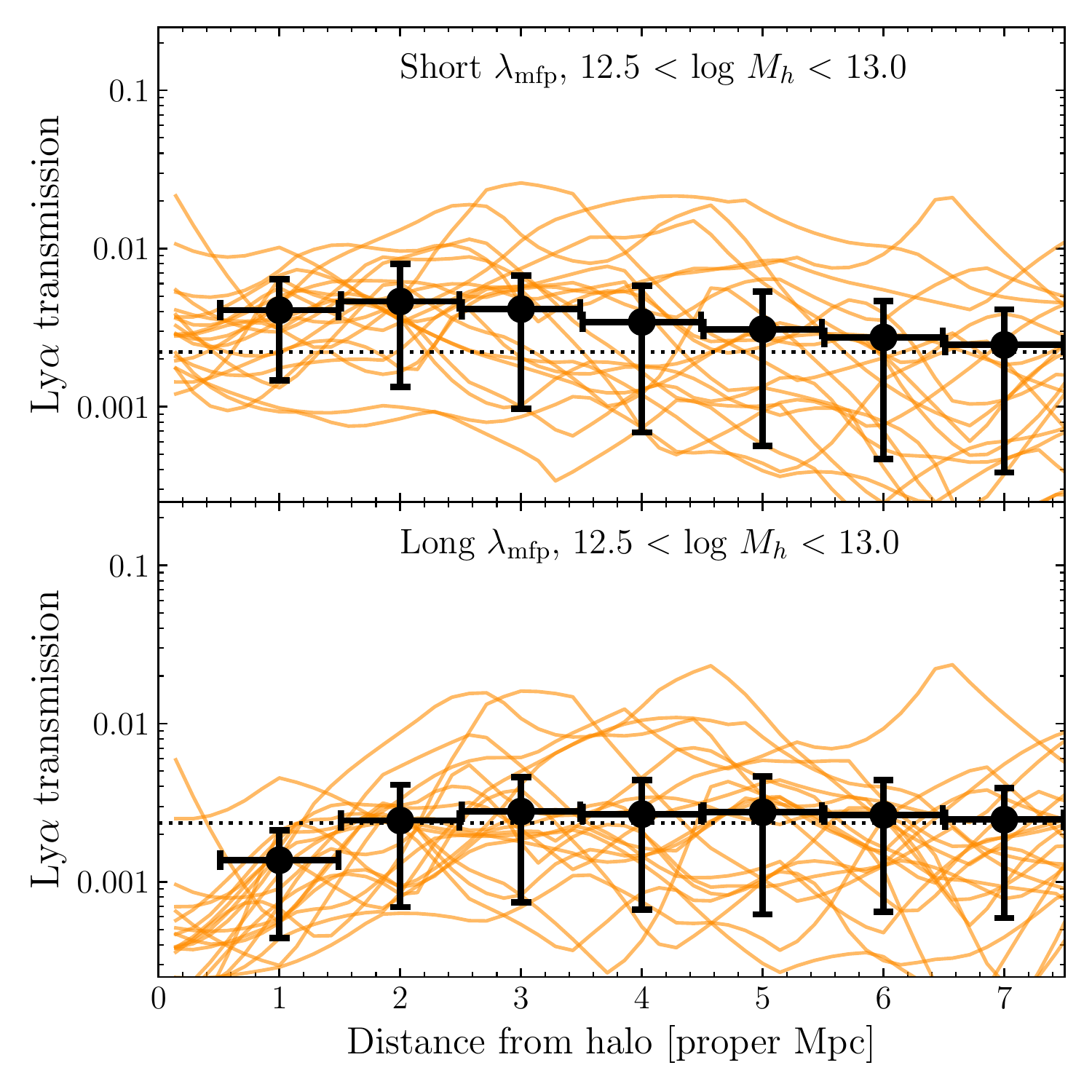}}
\end{center}
\caption{Similar to Figure~\ref{fig:tra_bias_var} but now including the effect of large-scale IGM density fluctuations on the Ly$\alpha$ optical depth along each line of sight.}
\label{fig:tra_bias_var2}
\end{figure}

To estimate the Ly$\alpha$ forest transmission resulting from the ionization rate profiles, we assumed that the effective optical depth of the Ly$\alpha$ forest (where $\tau_{\rm eff}\equiv-\ln{\langle F \rangle}$) behaves as $\tau_{\rm eff}=5.7(\Gamma_{\rm HI}/[2.5\times10^{-13}\,{\rm s}^{-1}])^{-0.55}$, as calibrated from the $z=6$ hydrodynamical simulation used by \citet{Davies19b}. In Figure~\ref{fig:tra_bias} we show the resulting mean Ly$\alpha$ forest transmission profiles for the same halo mass bins as Figure~\ref{fig:halo_bias}. The transmission profiles are qualitatively consistent with the predictions from \citetalias{Kakiichi18}, although a detailed comparison is made difficult due to the different method for populating halos with galaxies. A substantial excess in Ly$\alpha$ forest transmission is predicted at distances less than $\sim5$ proper Mpc from massive halos. 

As shown by the curves in Figure~\ref{fig:tra_bias_var}, however, there is considerable sightline-to-sightline variation in the expected (mean) transmission profile due to the stochasticity of ionizing background fluctuations. The black points and error bars in Figure~\ref{fig:tra_bias_var} represent the median and 16--84th percentile Ly$\alpha$ forest transmission of 5000 random skewers, neglecting any intrinsic scatter from small-scale IGM density fluctuations. While a conclusive detection of the biased background along the line of sight towards a single object seems unlikely, one could conceivably stack several lines of sight (i.e., the Ly$\alpha$ forest spectra of several quasars with proximate absorption) to improve the statistical significance.

The transmission profiles in Figures~\ref{fig:tra_bias} and \ref{fig:tra_bias_var} neglect the fact that the IGM nearby massive halos is overdense on large scales. To compute the density enhancement near massive halos, we interpolated through the Zel'dovich approximation density field binned onto the same $180^3$ grid as the ionizing background model. The resulting average overdensity profiles are shown in the top panel of Figure~\ref{fig:tra_bias2}. To approximate the effect of this overdensity on Ly$\alpha$ forest transmission, we assumed that the connection between the effective optical depth and density can be approximated via an analogy of the fitting formula in \citet{Davies19b} applied to density instead of $\Gamma_{\rm HI}$ through their connection to the number density of neutral hydrogen $n_{\rm HI}$. That is, $\tau_{\rm eff}\propto\Gamma_{\rm HI}^{-0.55}$ implies that $\tau_{\rm eff}\propto n_{\rm HI}^{0.55}$, and since $n_{\rm HI}\propto \Delta^{2-0.7(\gamma-1)}$ (e.g. \citealt{Weinberg97}), where $\gamma$ is the slope of the power-law temperature density relation in the IGM (e.g. \citealt{HG97}), we then have $\tau_{\rm eff}\propto\Delta^{0.55\times[2-0.7(\gamma-1)]}$. Assuming $\gamma=1.5$ as a fiducial value characteristic of the post-reionization IGM (e.g. \citealt{MUS16}) gives $\tau_{\rm eff}\propto\Delta^{0.9}$. Thus, to take density fluctuations into account, the effective optical depth was scaled by $\Delta^{0.9}$ along each skewer from the simulation\footnote{Note that this calculation is subtly different from simply applying the mean overdensity profile to the mean Ly$\alpha$ opacity profile due to correlations between density and ionization rate on the cell scale.}.

The lower two panels in Figure~\ref{fig:tra_bias2} show the resulting density-adjusted Ly$\alpha$ forest transmission profiles, and in Figure~\ref{fig:tra_bias_var2} we show the resulting sightline-to-sightline variations through the same skewers as Figure~\ref{fig:tra_bias_var}. In disagreement with Appendix B in \citetalias{Kakiichi18}, we find that the overdensity associated with massive halos eliminates the signal from the clustering-enhanced ionizing background unless the mean free path is short (i.e., unless the background fluctuations are very strong). This disagreement could be a result of the quasi-linear Zel'dovich approximation density field in the semi-numerical calculation versus the linear theory calculation in \citetalias{Kakiichi18}. The difference may also result from the way in which we modified the Ly$\alpha$ opacity as a function of $\Delta$. \citetalias{Kakiichi18} investigated modifying the Ly$\alpha$ \emph{transmission} directly, instead of the effective optical depth, assuming a fiducial linear bias of $b_\alpha=-1$. In comparison, modifying the effective optical depth directly behaves non-linearly in transmitted flux space, with $\delta F = \bar{F}^{\delta \tau}-1$. 

Nevertheless, even in the presence of the large-scale overdensity associated with massive halos, in the short $\lambda_{\rm mfp}$ model a modest excess in the transmitted Ly$\alpha$ forest flux persists at large distances from the quasar host halo, whose magnitude depends on the halo mass. However, the effect is quite weak, and care should be taken to ensure that the signal cannot be tampered with by the quasar -- even indirectly. As shown in the next section, this fear may very well be realized.

\section{The ``ghost" proximity effect} \label{sec:ghost}

While an optically-thick proximate absorber will block the vast majority of the ionizing radiation from the quasar along the line of sight, the rest of the surrounding IGM out to several proper Mpc will nevertheless have a greatly enhanced ionizing background, because the absorber only covers a modest solid angle of the quasar's emission. This elevated ionizing background will increase the local mean free path of ionizing photons\footnote{As noted by \citet{D'Aloisio18}, this also implies that measurements of the mean free path from stacking non-DLA quasars (e.g. \citealt{Worseck14}) could be biased high by a factor of two relative to typical IGM environments at $z\sim5$.}, and this could lead to an increase in the ionizing background along the line of sight \emph{indirectly} due to the enhanced contribution from distant sources. Thus a ``ghost" of the proximity effect may be present, and could in principle be confused with the signal from the clustering of nearby galaxies. 

\begin{figure}
\begin{center}
\resizebox{8.5cm}{!}{\includegraphics[trim={6em 1em 9em 0em},clip]{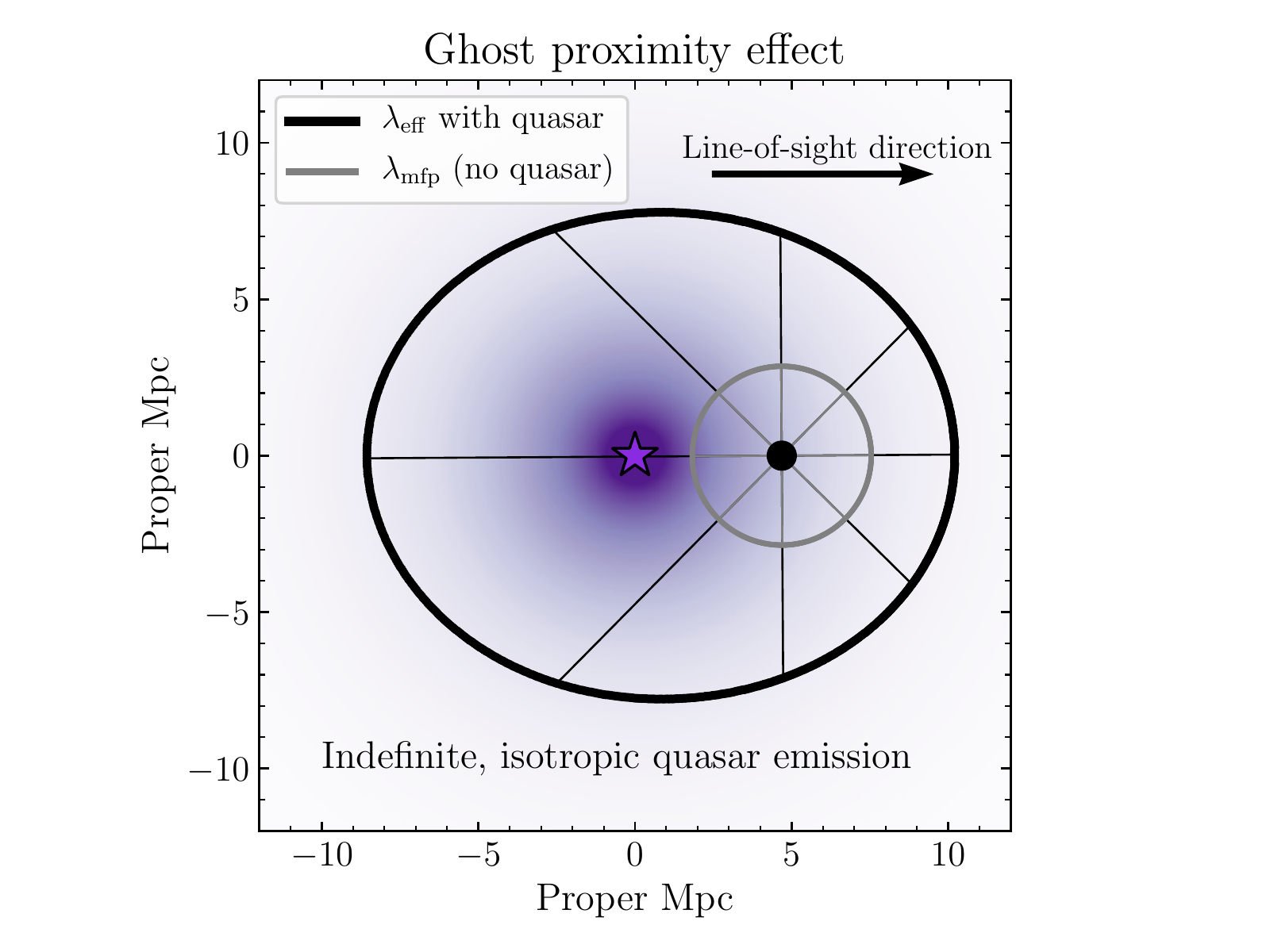}}
\end{center}
\caption{Schematic diagram of the ghost proximity effect. The quasar is located at the origin, represented by the purple star, and the purple shading represents the ionizing radiation emitted by the quasar. A particular location along the line of sight (horizontal axis) is highlighted by the black dot. The grey circle around this location has a radius equal to $\lambda_{\rm mfp}$, while the black oval shows the effective mean free path in the presence of the quasar radiation, i.e. the radial distance from the black dot to the point where the integrated optical depth is equal to one (equation~\ref{eqn:lam_eff}).}
\label{fig:gpe_basic}
\end{figure}

\begin{figure}
\begin{center}
\resizebox{8.5cm}{!}{\includegraphics[trim={1em 1em 1em 1em},clip]{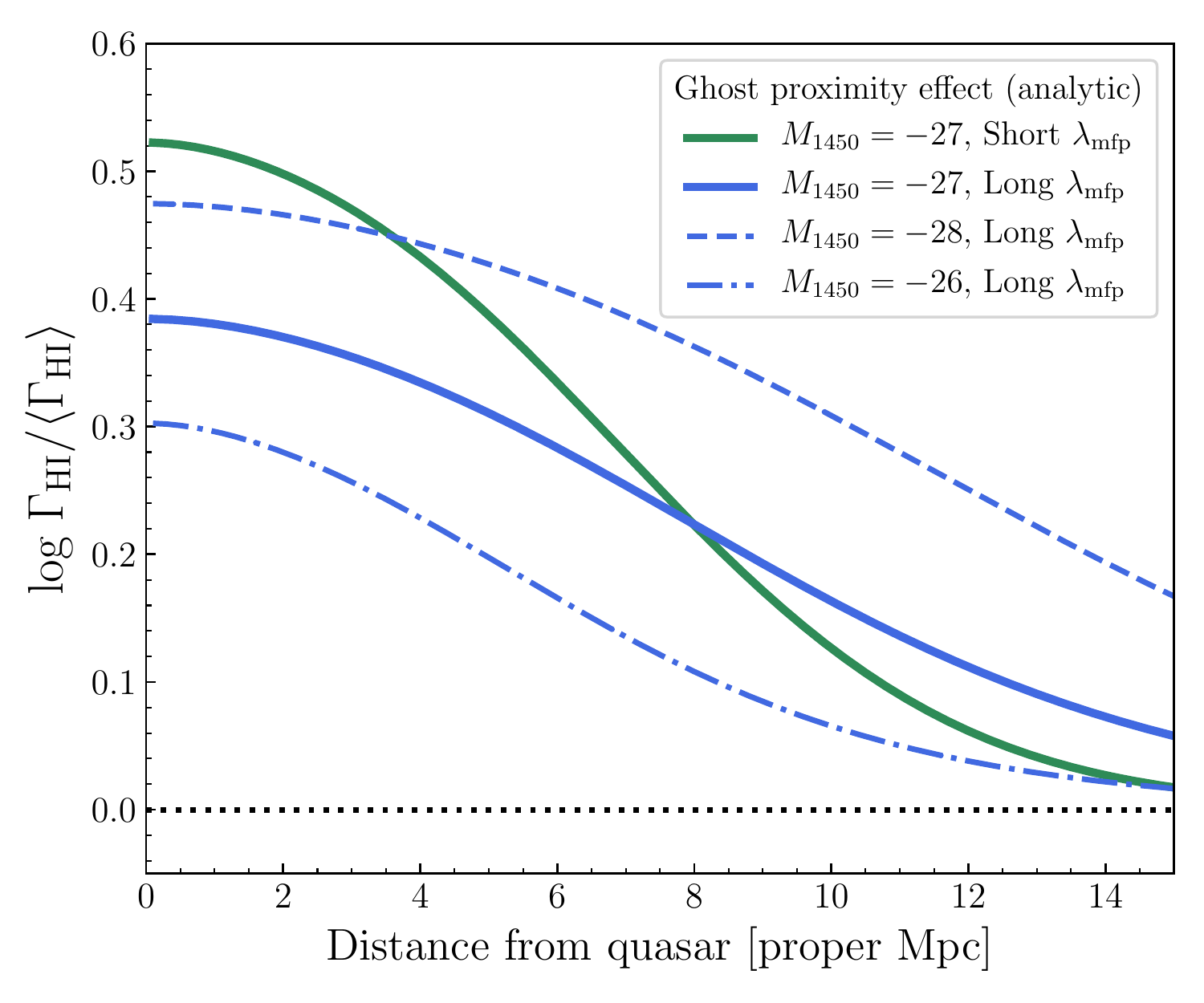}}
\end{center}
\caption{Analytic model for the ghost proximity effect. The green and blue curves correspond to a $M_{1450}=-27$ quasar in the short $\lambda_{\rm mfp}$ and long $\lambda_{\rm mfp}$ cases, respectively. The dashed and dot-dashed curves show the effect of increasing and decreasing the luminosity of the quasar by one magnitude. The horizontal dotted line indicates the mean background ionization rate.}
\label{fig:analytic_ghost}
\end{figure}

\begin{figure*}
\begin{center}
\resizebox{18cm}{!}{\includegraphics[trim={7em 4em 1em 7em},clip]{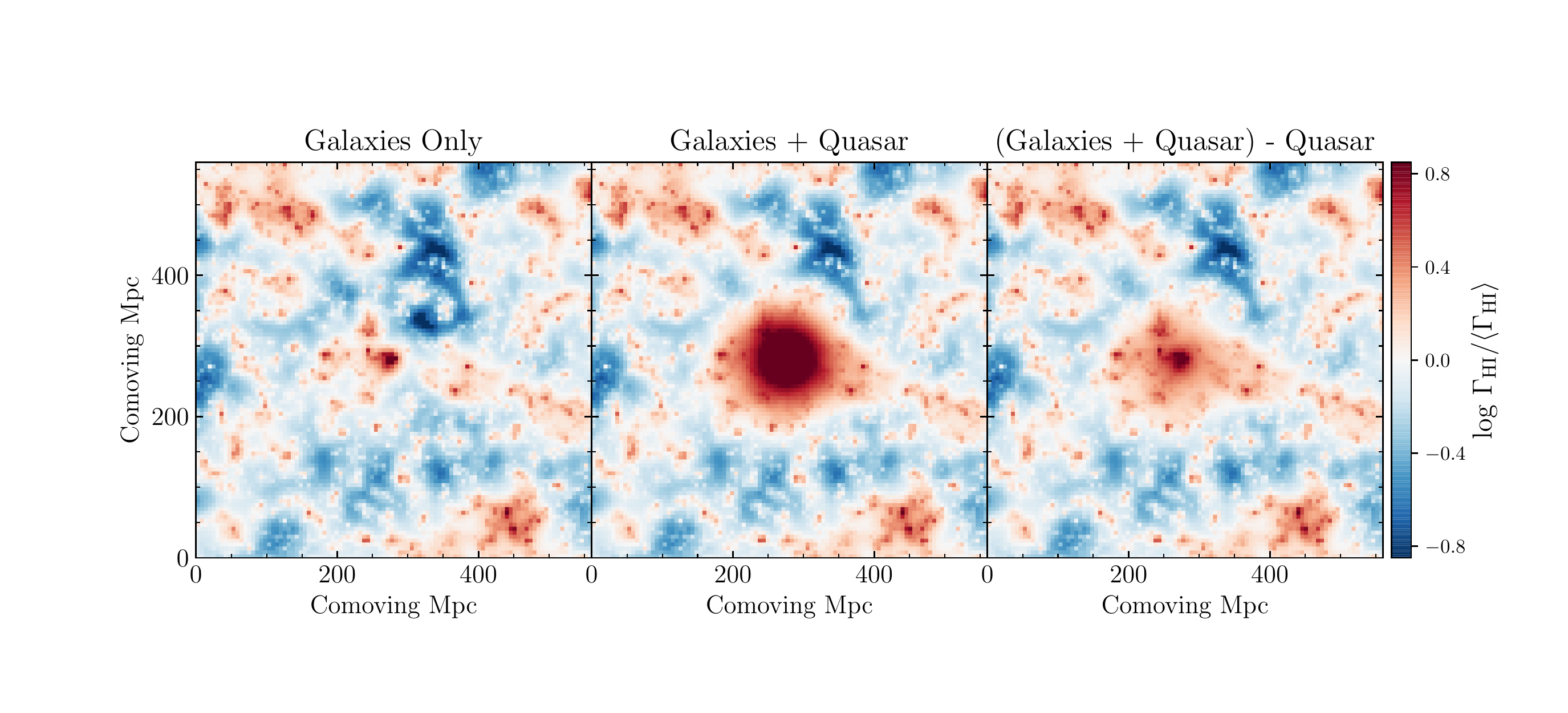}}
\end{center}
\caption{Cropped slices ($570\,\times\,570$ comoving Mpc$^2$) through the short $\lambda_{\rm mfp}$ fluctuating ionizing background model (left), the model with an added $M_{1450}=-27$ quasar (middle), and the model including the quasar but with the direct contribution from the quasar subtracted out (right). The difference between the left and right panels thus demonstrates the impact of the ``ghost" proximity effect. The slices are centered on the most massive dark matter halo in the semi-numerical box, with $M_{\rm h} = 10^{13}\,{\rm M}_\odot$.}
\label{fig:uvb_ghost}
\end{figure*}

\begin{figure}
\begin{center}
\resizebox{8.5cm}{!}{\includegraphics[trim={1em 1em 1em 1em},clip]{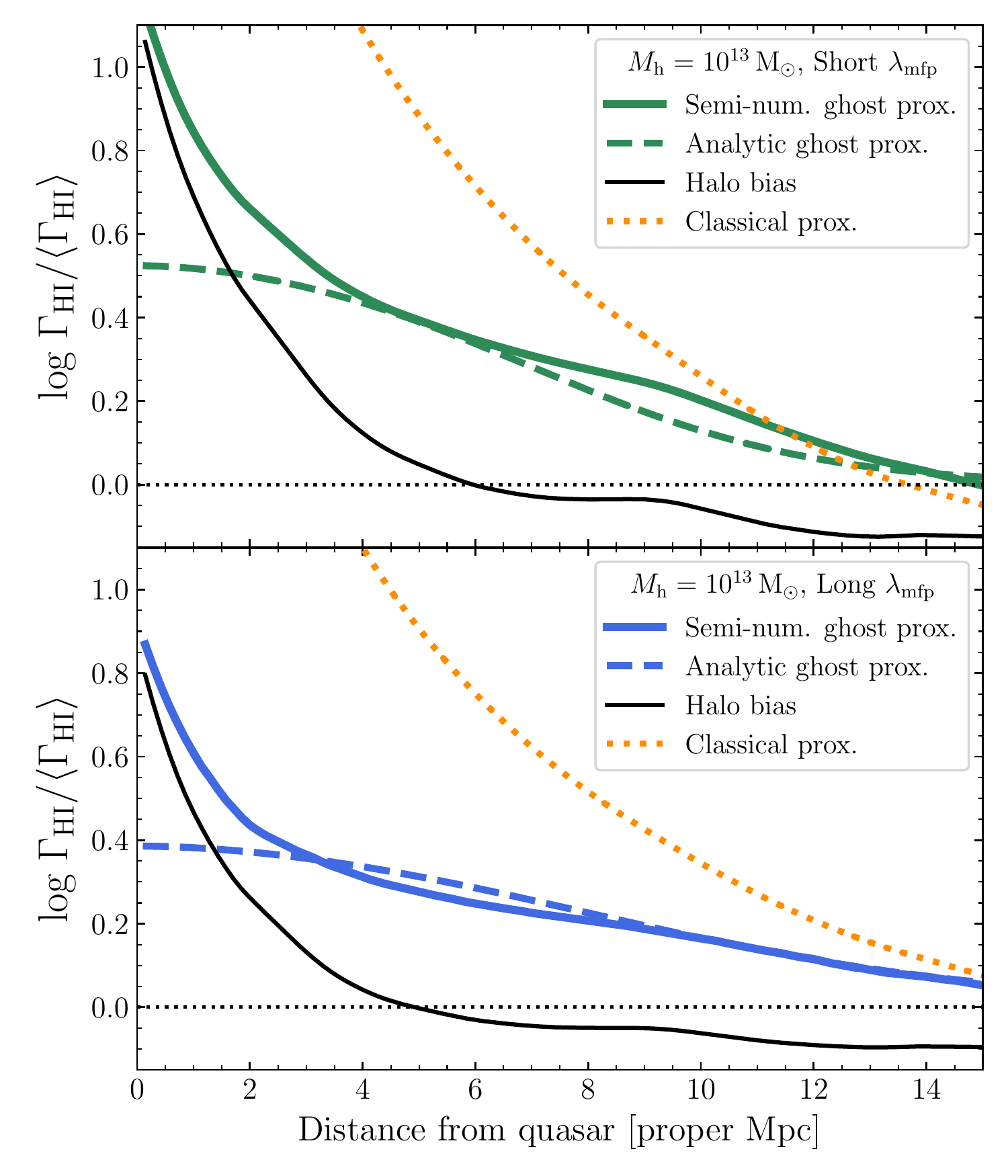}}
\end{center}
\caption{Ghost proximity effect in the semi-numerical ionizing background simulations. The top and bottom panels show the short $\lambda_{\rm mfp}$ and long $\lambda_{\rm mfp}$ models, respectively. The solid green and blue curves show the average excess photoionization rate from the semi-numerical simulations where a $M_{1450}=-27$ quasar was placed in the most massive halo ($M_{\rm h}=10^{13}\,{\rm M}_\odot$), while the dashed green and blue curves show the analytic predictions from Figure~\ref{fig:analytic_ghost}.
The dotted orange curves show the direct contribution from the quasar that would be present without a proximate absorber.}
\label{fig:seminum_ghost}
\end{figure}

\begin{figure}
\begin{center}
\resizebox{8.5cm}{!}{\includegraphics[trim={1em 1em 1em 1em},clip]{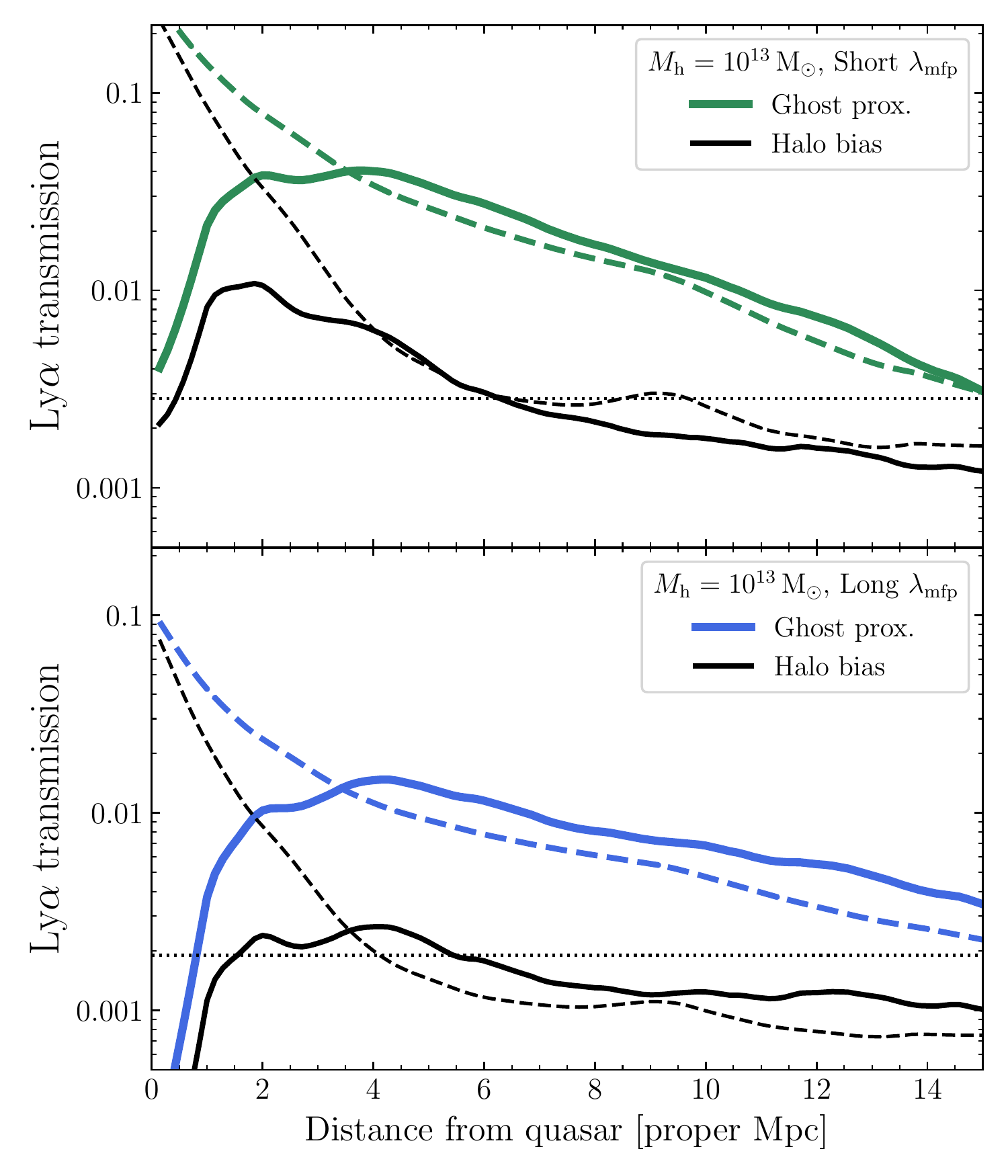}}
\end{center}
\caption{Average Ly$\alpha$ forest transmission profiles from the semi-numerical ghost proximity effect around the most massive halo in the simulation. The thick green and blue curves show the ghost proximity effect signal for the short $\lambda_{\rm mfp}$ and long $\lambda_{\rm mfp}$ models, respectively, while the thin black curves show the corresponding signal from the halo bias alone. Solid curves include the effect of density fluctuations on the Ly$\alpha$ opacity while the dashed curves ignore it. The horizontal dotted line indicates the mean Ly$\alpha$ forest transmission.}
\label{fig:seminum_ghost_tra}
\end{figure}

\subsection{Analytic ghost proximity effect} \label{sec:analytic_ghost}

The expected order of magnitude of the ghost proximity effect can be estimated via a simple analytic approach. For simplicity, we assume that the quasar emits ionizing photons isotropically and has been shining indefinitely, thus illuminating all of its surroundings with the exception of the pencil beam along the line of sight which is blocked by the DLA. This quasar-illuminated IGM will have a longer mean free path, and because $\Gamma_{\rm HI}\propto\lambda_{\rm mfp}$ (in the absorption-limited regime, \citealt{MW03}), the ionization rate will also increase.

Figure~\ref{fig:gpe_basic} illustrates the basic picture. The quasar is shown as the purple star, and a location along the line of sight is shown by the black dot. The grey circle shows $\lambda_{\rm mfp}$ around this location -- i.e., the typical volume from which an average region in the IGM would receive ionizing photons. The black oval shows the ``effective" mean free path $\lambda_{\rm mfp}$, defined as
\begin{equation} \label{eqn:lam_eff}
\int_0^{\lambda_{\rm eff}} \frac{d\tau_{\rm HI}}{dx} dx = 1,
\end{equation}
where $x$ is the distance along a trajectory starting at the location along the line of sight (e.g. the thin black lines in Figure~\ref{fig:gpe_basic}), and $d\tau_{\rm HI}/dx$ is the IGM opacity to ionizing photons given by
\begin{equation}
\frac{d\tau_{\rm HI}}{dr} = \lambda_{\rm mfp}^{-1} \left(\frac{\Gamma_{\rm HI,tot}}{\Gamma_{\rm b}}\right)^{-2/3}
\end{equation}
following the dependence assumed for the fluctuating ionizing background simulations in the previous section (although neglecting the density dependence for simplicity), where $\Gamma_{\rm HI,tot}=\Gamma_{\rm q}+\Gamma_{\rm b}$ is the total ionization rate and $\Gamma_{\rm b}=2.0\times10^{-13}$ s$^{-1}$ is the mean ionization rate in the IGM. The ionizing radiation from the quasar is attenuated by the IGM out to the distance $r$ as
\begin{equation}
\Gamma_{\rm q}(r) = \Gamma_{{\rm q},0} \left(\frac{r}{r_0}\right)^{-2} e^{-\int_0^r \frac{d\tau_{\rm HI}}{dr'} dr'},
\end{equation}
where $\Gamma_{{\rm q},0}$ is the unattenuated photoionization rate due to the quasar at a distance $r_0$ and $d\tau_{\rm HI}/dr$ represents the IGM opacity as a function of distance from the quasar. Assuming the quasar SED from \citet{Lusso15}, a quasar with $M_{1450}=-27$ (i.e. the absolute magnitude at rest-frame $1450$\,\AA) has $\Gamma_{{\rm q},0}\approx 3.2\times10^{-11}$ s$^{-1}$ at $r_0=1$ proper Mpc.

An initial estimate of the strength of the ghost proximity effect can then be obtained by multiplying the background photoionization rate by a factor $\lambda_{\rm eff}/\lambda_{\rm mfp}$ at a series of positions along the line of sight. The angular dependence of $\lambda_{\rm eff}$ is non-trivial, as shown in Figure~\ref{fig:gpe_basic}, so for simplicity we assume that $\lambda_{\rm eff}$ in the transverse direction (i.e. the vertical axis in Figure~\ref{fig:gpe_basic}) acts as a representative average. In Figure~\ref{fig:analytic_ghost} we show the resulting predictions for the strength of the ghost proximity effect as a function of $\lambda_{\rm mfp}$ and the quasar absolute magnitude. The enhancement is quite strong, amounting to a factor of $\sim2.5$ at 5 proper Mpc, and it is stronger than the bias due to nearby galaxies (Figure~\ref{fig:halo_bias}) everywhere except for $R<1$ proper Mpc around the most massive halos in Figure~\ref{fig:halo_bias}. Perhaps more interestingly, the predicted ghost proximity effect extends out to very large distances, $>10$ proper Mpc, suggesting that it may be a much easier signal to detect. 

\subsection{Semi-numerical ghost proximity effect} \label{sec:seminum_ghost}

To quantify the ghost proximity effect in 3D, we re-computed the semi-numerical fluctuating background models from \S~\ref{sec:uvb} several times with a single luminous quasar ($M_{1450}=-27$) at the location of different massive ($>10^{12.5}\,{\rm M}_\odot$) dark matter halos. As in the analytic model, we assumed indefinite, isotropic quasar emission. In Figure~\ref{fig:uvb_ghost} we show slices from one of these fluctuating ionizing background models centered on the most massive halo in the short $\lambda_{\rm mfp}$ simulation with $M_{\rm h} = 10^{13}\,{\rm M}_\odot$. From left to right, the panels show slices from the original simulation, the simulation with a quasar added at the location of the halo, and then the latter simulation but with the \emph{direct} contribution from the quasar removed. The excess ionization rate in the right panel compared to the left panel demonstrates the ghost proximity effect -- an enhancement of the ionizing background without any direct contribution of ionizing photons from the quasar. As expected from the analytic analysis in \S~\ref{sec:analytic_ghost} above, the magnitude of the ghost proximity effect is quite strong, and the effect dominates over the underlying background fluctuations out to very large distances.

\begin{figure}
\begin{center}
\resizebox{8.5cm}{!}{\includegraphics[trim={1em 1em 1em 1em},clip]{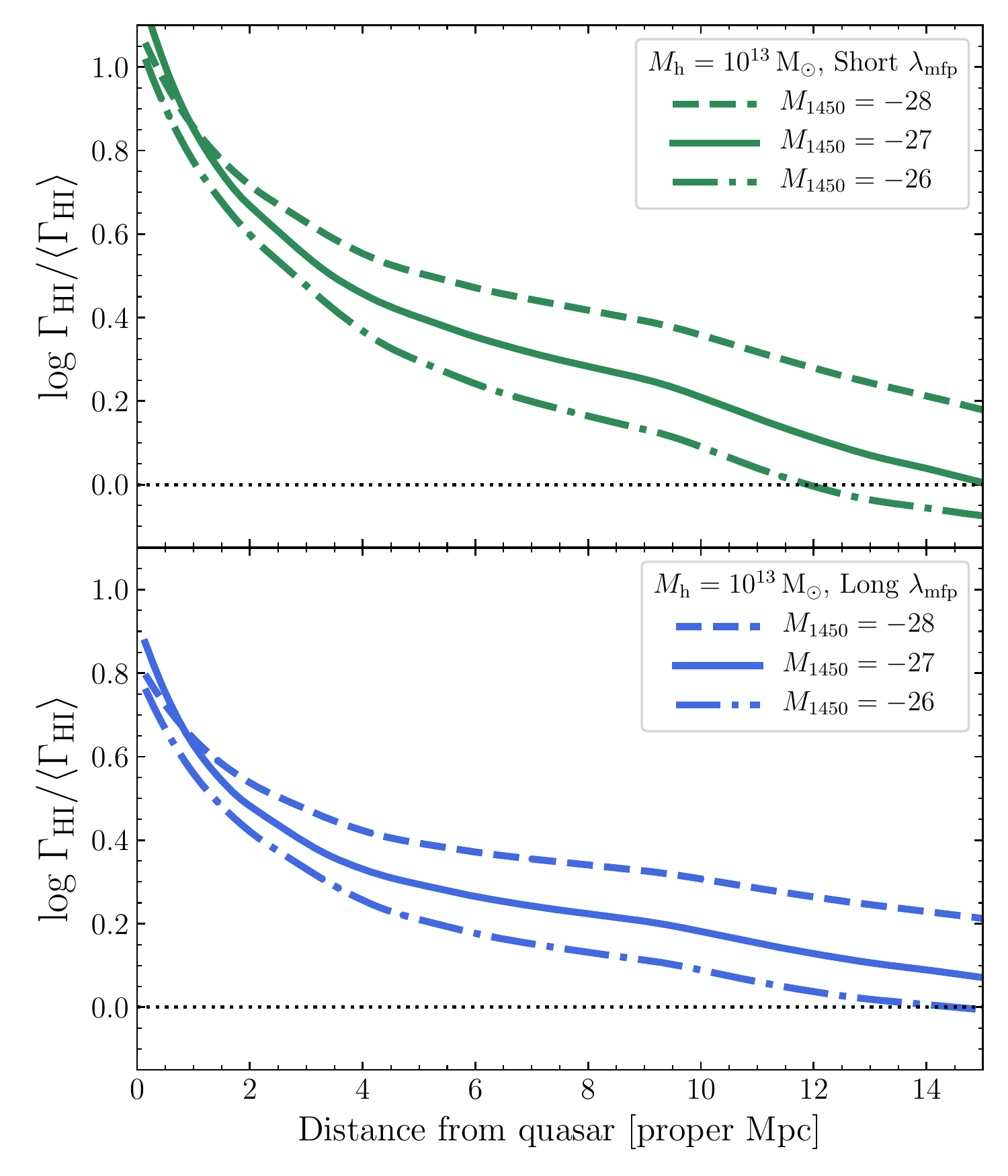}}
\end{center}
\caption{Similar to Figure~\ref{fig:seminum_ghost} but now showing the dependence of the mean ghost proximity profile on quasar luminosity, where the dot-dashed, solid, and dashed curves show the strength of the ghost proximity effect for quasars with $M_{1450}=-26$, $-27$, and $-28$, respectively.}
\label{fig:seminum_ghost2}
\end{figure}

\begin{figure}
\begin{center}
\resizebox{8.5cm}{!}{\includegraphics[trim={1em 1em 1em 1em},clip]{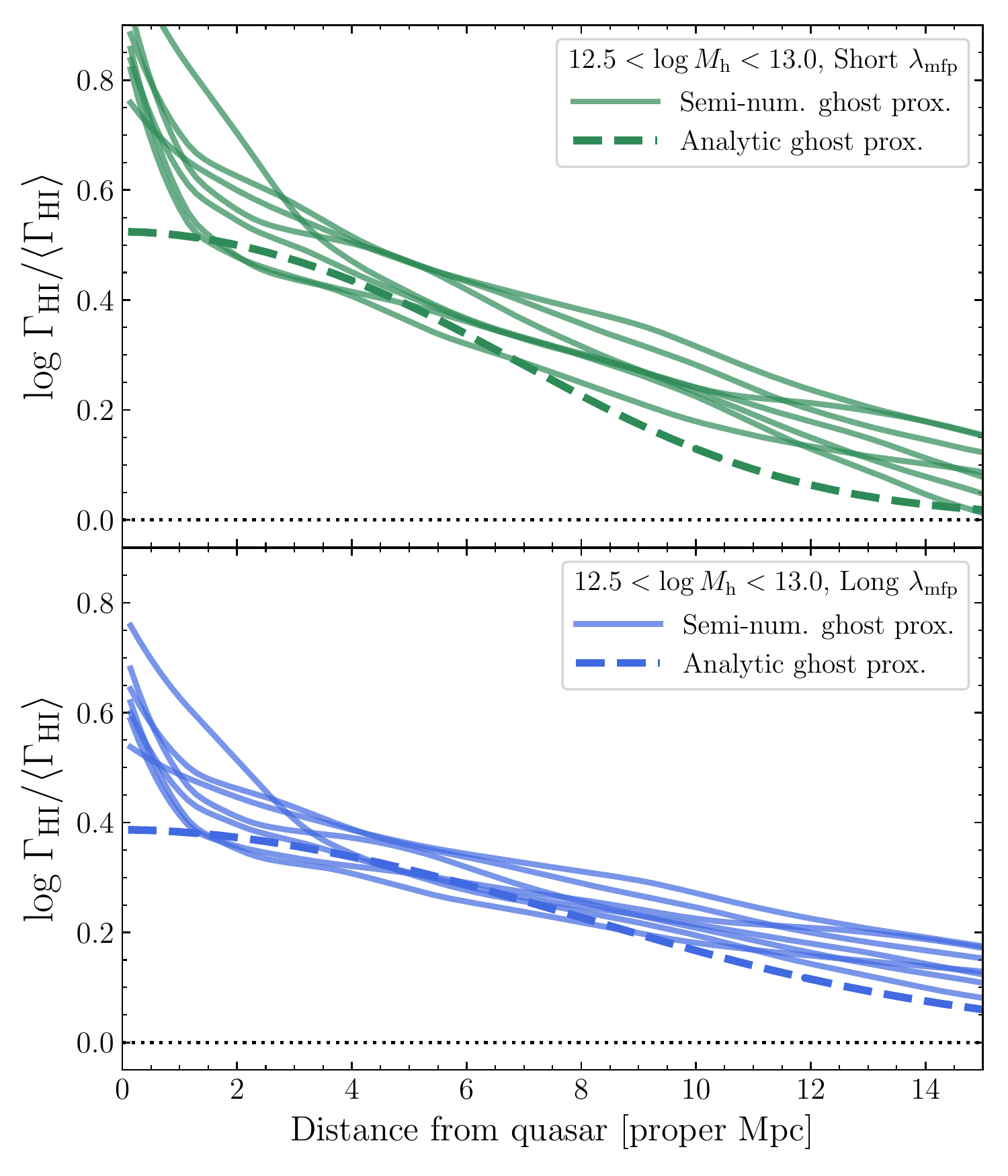}}
\end{center}
\caption{Similar to Figure~\ref{fig:seminum_ghost} but now showing mean ghost proximity effect profiles (blue/green) and halo bias (black) for ionizing background simulations where the quasar was placed in seven different halos with $12.5 < \log{M_{\rm h}} < 13.0$. Dashed curves show the corresponding analytic model predictions for the ghost proximity effect.}
\label{fig:seminum_ghost3}
\end{figure}

In Figure~\ref{fig:seminum_ghost} we show the radially-averaged profile of the photoionization rate in the short $\lambda_{\rm mfp}$ (top) and long $\lambda_{\rm mfp}$ (bottom) models, centered on the massive quasar-hosting halo in Figure~\ref{fig:uvb_ghost}, compared to the analytic prediction. The two agree surprisingly well given the simplicity of the analytic calculation. 
At distances $\lesssim2$ proper Mpc from the quasar, the effect of halo bias (not included in the analytic model) is still evident, but at large distances the signal appears to be completely dominated by the ghost proximity effect. The dotted orange curves in Figure~\ref{fig:seminum_ghost} show the expected photoionization rate profile along the line of sight to an absorber-free quasar (i.e. with the direct contribution from the quasar included), but \emph{without} the ghost proximity effect, i.e. assuming that the quasar has no impact on the contribution from surrounding galaxies. Interestingly, the mean free path boost that leads to the ghost proximity effect should have a non-negligible impact on sightlines without proximate absorption systems as well, substantially extending the region along the line of sight in which the ionizing background is ``contaminated" by quasar radiation. While this likely has little effect on measurements of proximity zone sizes via their traditional definition where the quasar light alone dominates (e.g. \citealt{Fan06,Carilli10,Eilers17,Davies19b}), it motivates very large exclusion regions blueward of rest-frame Ly$\alpha$ when estimating the optical depth of the Ly$\alpha$ forest (i.e., in the general IGM) from quasar spectra \citep{Eilers18,Bosman18}. 

In Figure~\ref{fig:seminum_ghost_tra} we show the mean Ly$\alpha$ forest transmission signal from the ghost proximity effect in the semi-numerical simulations compared to the halo bias signal. The ghost proximity effect in Ly$\alpha$ forest transmission is very strong, amounting to a factor of $\sim10$ ($\sim5$) boost for the short (long) $\lambda_{\rm mfp}$ models at a distance of $\sim5$ proper Mpc from the quasar, with a non-negligible effect even at distances of $\ga10$ proper Mpc. The ghost proximity effect clearly dominates over the signal from halo bias, although detailed modelling of the increased signal at distances less than a few proper Mpc may allow both effects to be disentangled.

An additional suite of ionizing background simulations were run on a lower resolution grid, $144^3$, to explore the dependence on quasar luminosity and the scatter of the ghost proximity effect between different halo environments. In Figure~\ref{fig:seminum_ghost2} we show the ghost proximity effect photoionization rate profiles for quasars with $M_{1450}=-26$, $-27$, and $-28$ located in the same $10^{13}\,{\rm M}_\odot$ halo, showing that the dependence on quasar luminosity is very similar to the analytic prediction in Figure~\ref{fig:analytic_ghost}. In Figure~\ref{fig:seminum_ghost3} we show the ghost proximity effect for seven different halos in the mass range $12.5 < \log{M_{\rm h}} < 13.0$, demonstrating the moderate scatter in the mean profiles between different halo environments.

\subsection{Revisiting the assumption of indefinite, isotropic emission} \label{sec:causal}

The question remains whether this ghost proximity effect truly exists in nature, or if it merely represents the result of a series of incorrect assumptions. Notably, we assumed that the quasar emitted ionizing photons isotropically and indefinitely, which is unlikely to represent the true emission pattern or activity duration. 

\begin{figure}
\begin{center}
\resizebox{8.5cm}{!}{\includegraphics[trim={6em 1em 9em 0em},clip]{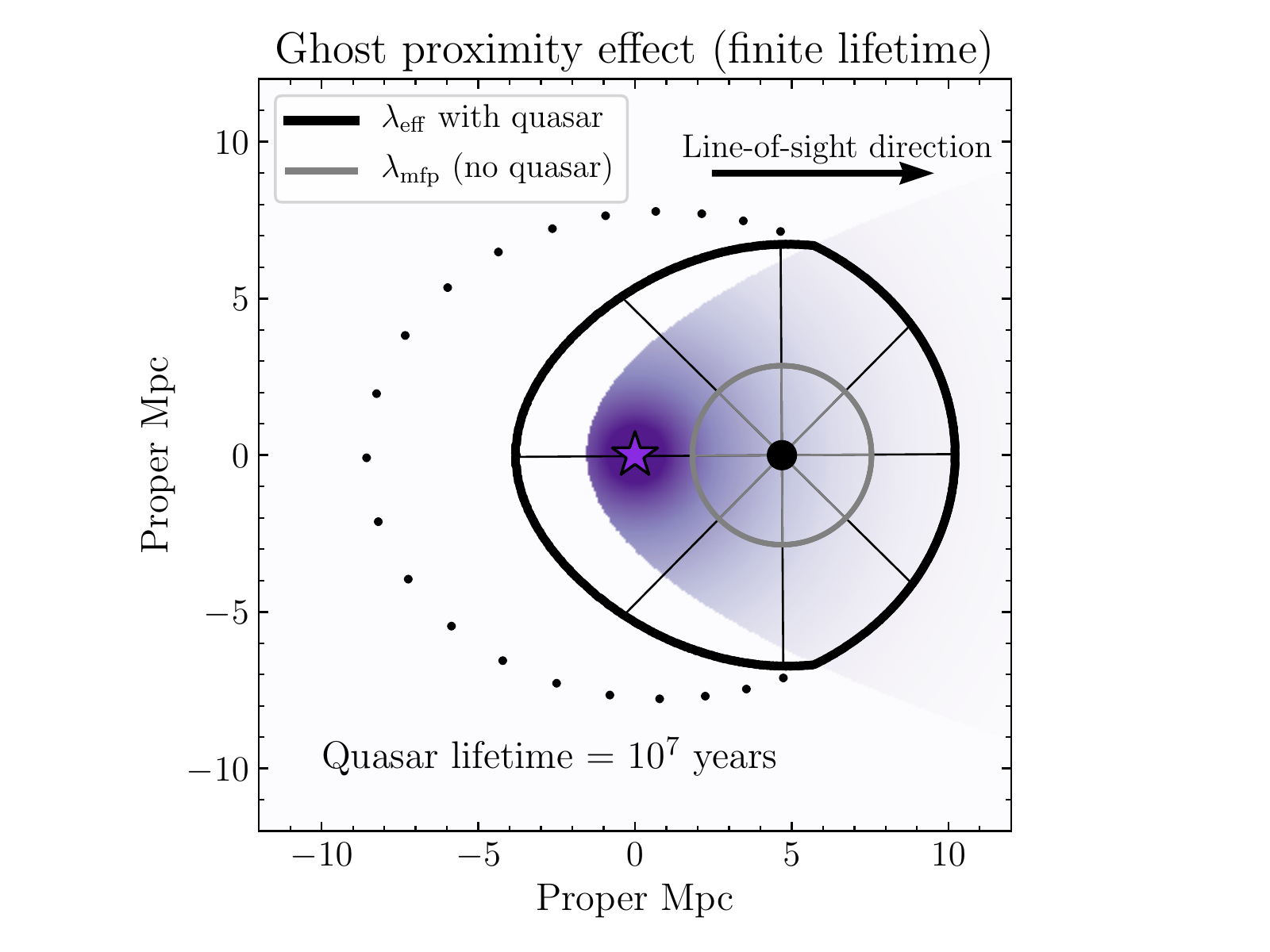}}
\end{center}
\caption{Similar to Figure~\ref{fig:gpe_basic} but now highlighting the effect of a finite quasar lifetime on the ghost proximity effect. The black dotted oval shows the original effective mean free path assuming an indefinite quasar lifetime, while the black solid oval shows the new effective mean free path due to the smaller illuminated volume along the light cone assuming a quasar lifetime of $10^7$ years (blue shaded region).}
\label{fig:gpe_tq}
\end{figure}

\begin{figure}
\begin{center}
\resizebox{8.5cm}{!}{\includegraphics[trim={6em 1em 9em 0em},clip]{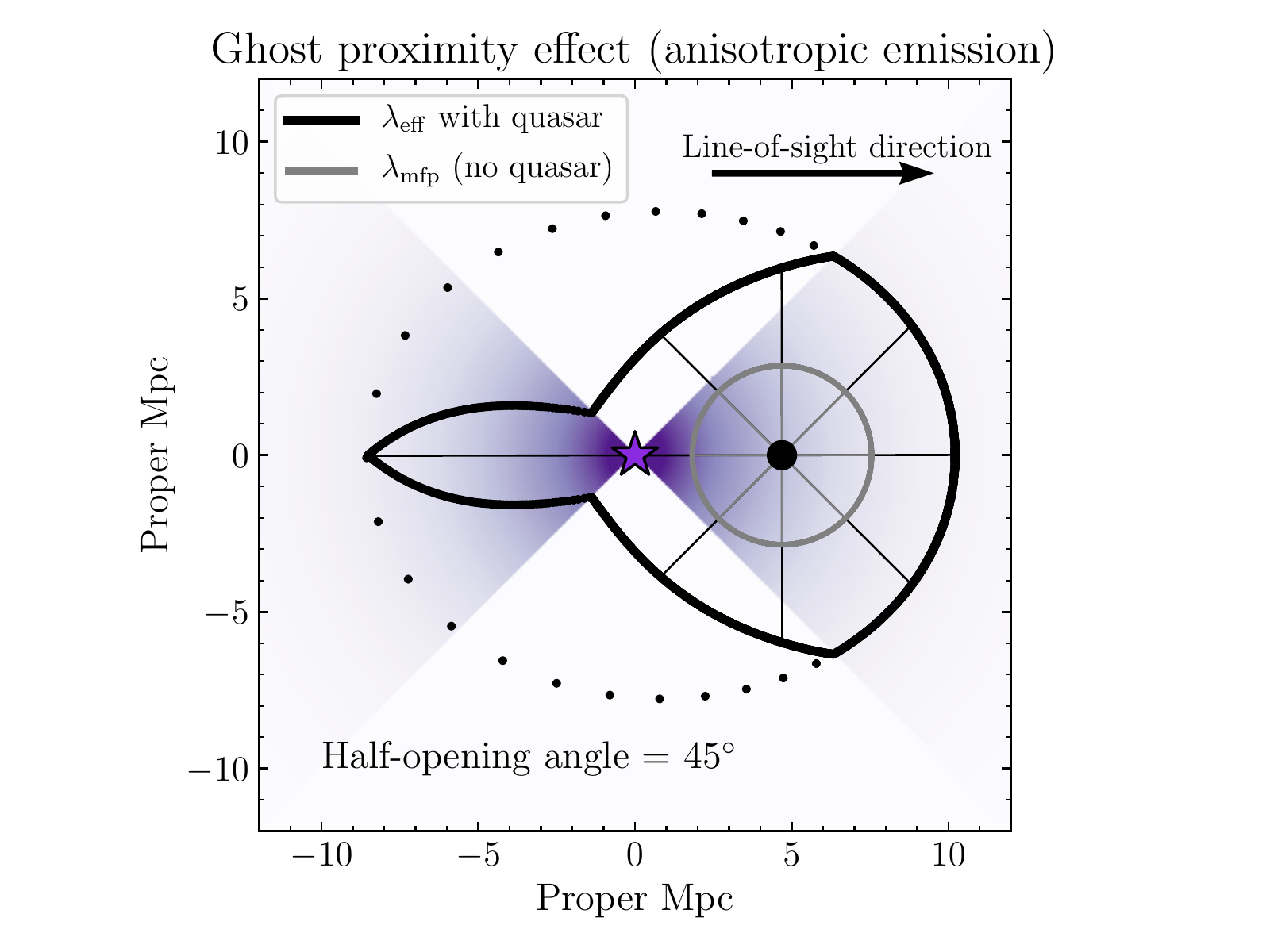}}
\end{center}
\caption{Similar to Figure~\ref{fig:gpe_basic} and Figure~\ref{fig:gpe_tq} but now highlighting the effect of anisotropic emission, here chosen to be a biconical shape directed along the line of sight with a half-opening angle of $45^\circ$ (blue shaded region).}
\label{fig:gpe_theta}
\end{figure}

If one considers a ``lightbulb" quasar lightcurve, i.e. the quasar turned on at some point in the past and has been shining steadily since then, the maximum volume that is causally connected to the quasar emission will be represented by a paraboloid with the quasar at the focus (e.g. \citealt{VC08}). If the transverse extent of this paraboloid, given approximately by $r_\perp \approx \sqrt{(ct_{\rm q}-r_\parallel)^2-r_\parallel^2}$ where $t_{\rm q}$ is the age of the quasar \citep{Schmidt18echo}, is much smaller than the enhanced mean free path, then the ghost proximity effect would be much weaker. In Figure~\ref{fig:gpe_tq} we illustrate the effect of a $10^7$ year quasar lifetime on the angular distribution of the effective mean free path (e.g. Figure~\ref{fig:gpe_basic}). Comparing the finite lifetime volume encompassed by the effective mean free path (solid black) to the indefinite lifetime volume (dotted black) suggests that the ghost proximity effect would indeed be weakened. Similarly, any anisotropy in the quasar emission pattern (as implied by the unification model, e.g. \citealt{Antonucci93,UP95}) would also reduce the quasar-illuminated volume. In Figure~\ref{fig:gpe_theta} we show the effect of biconical emission with a half-opening angle of $45^\circ$ on the effective mean free path, analogously to Figure~\ref{fig:gpe_tq}, which similarly shows a substantial decrease in the effective mean free path relative to the isotropic case. 

Measurements of the ghost proximity effect could then be used to constrain these fundamental parameters of the quasar phenomenon by indirectly measuring the strength of the \emph{transverse} proximity effect (e.g. \citealt{FL11,Schmidt17a,Schmidt18}). We leave a more detailed exploration of 3D calculations of the effects of varying anisotropy and quasar activity on the ghost proximity effect to future work. In addition to constraining the properties of quasar emission, a conclusive detection of this effect would also prove the existence of a strong ionizing background dependence of the mean free path, indicating that measurements of the mean free path along the line of sight to $z\ga5$ quasars \citep{Worseck14} should account for this effect\footnote{Note that the mean free path along the line of sight is \emph{not} sensitive to the geometry or timescale of quasar emission, so any single conclusive detection of the ghost proximity effect (presumably towards a particularly long-lived, relatively isotropically-emitting quasar) would heavily imply that the stacking measurements are biased.} \citep{D'Aloisio18}.

\section{Observational evidence for the ghost proximity effect: PSO J056--16} \label{sec:j056}

\begin{figure}
\begin{center}
\resizebox{8.5cm}{!}{\includegraphics[trim={1em 1em 1em 1em},clip]{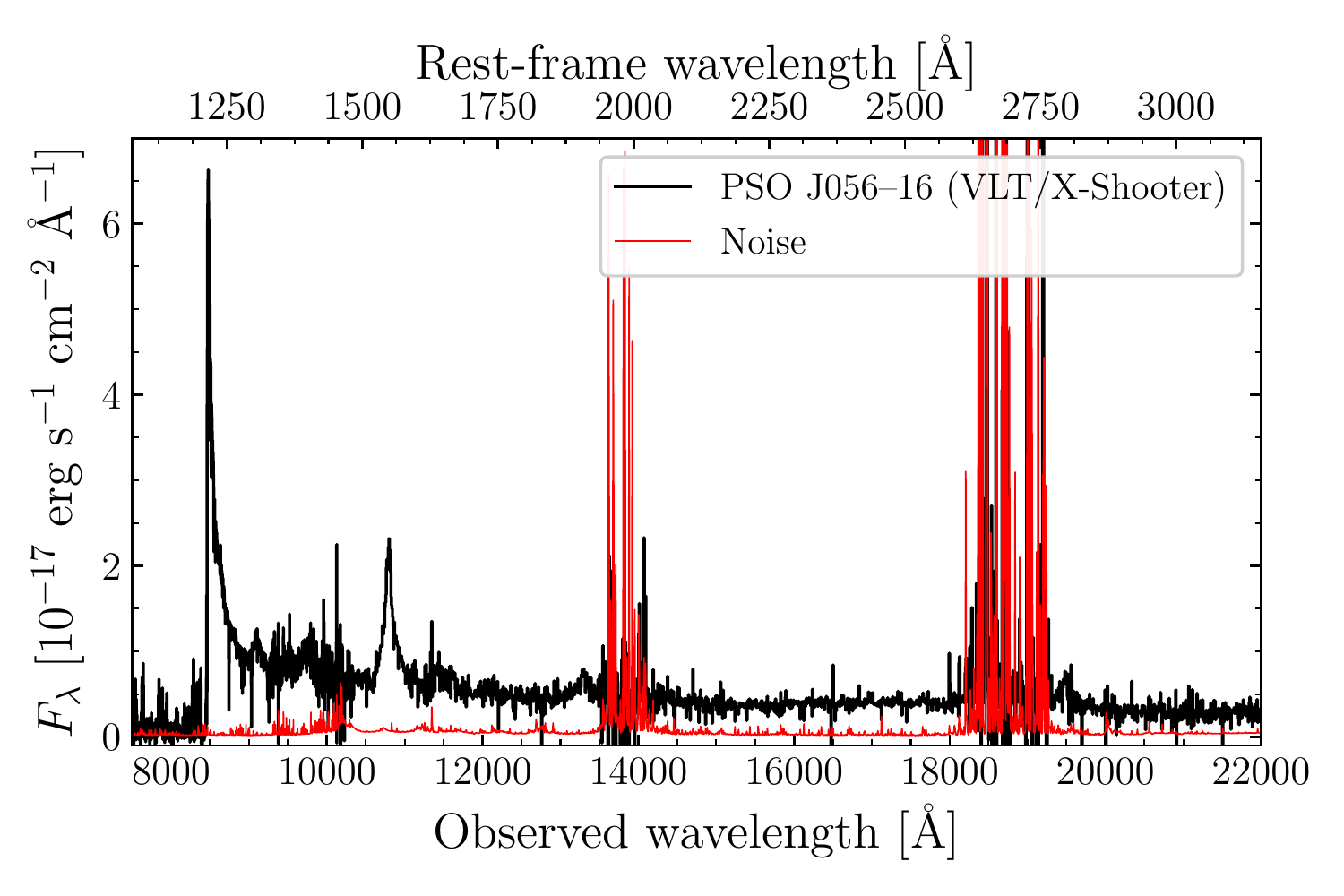}}
\end{center}
\caption{Combined VLT/X-Shooter VIS+NIR spectrum of PSO J056--16 after median filtering with an 11-pixel window (black curve), and its $1\sigma$ noise vector (red curve).}
\label{fig:j056_full}
\end{figure}

In Figure~\ref{fig:j056_full}, we show a preliminary spectrum of the $z\sim6$, $M_{1450}\sim-27$ quasar PSO J056--16 \citep{Banados16}, observed for 80 minutes (2 observation blocks) with VLT/X-Shooter (PI: Farina). As discussed below, this quasar appears to have a proximate DLA system. The detailed properties of the J056--16 spectrum and its proximate DLA will be discussed in future work (Farina et al., in prep.), but we summarize here the procedure used to independently reduce and analyze the spectrum in the context of this work. The raw data and associated calibration data were acquired from the public ESO archive and reduced with the ESO X-Shooter pipeline (version 2.9.3; \citealt{Modigliani10}). The output 1D spectra from each observation block were then stacked via an inverse variance-weighted average.
Finally, telluric absorption was corrected using an absorption model derived with \texttt{molecfit} \citep{Smette15} on a telluric standard star observed during the same night. 

\begin{figure}
\begin{center}
\resizebox{8.5cm}{!}{\includegraphics[trim={1em 1em 1em 1em},clip]{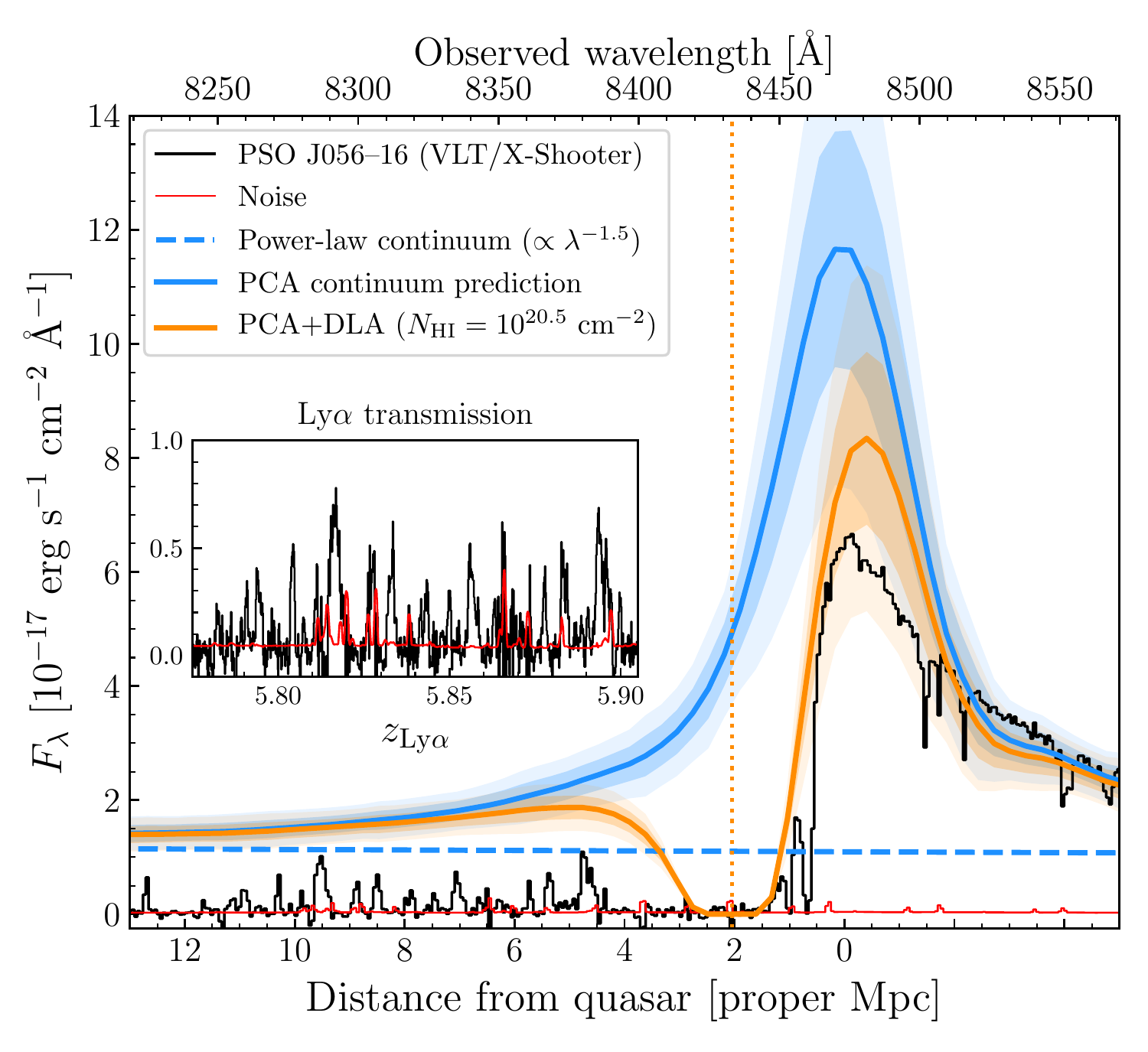}}
\end{center}
\caption{VLT/X-Shooter spectrum of PSO J056--16 re-binned to roughly 1 resolution element per pixel (black curve), and its associated $1\sigma$ noise vector (red curve). The dashed blue curve shows a power law continuum model $\propto\lambda^{-1.5}$ extrapolating from $\lambda_{\rm rest}=1280$\,\AA. The blue curve shows a PCA reconstruction of the quasar spectrum close to Ly$\alpha$ using the information contained in the $\lambda_{\rm rest}=1280$--$2850$\,\AA\ region, while the blue shaded regions show the associated 1 and $2\sigma$ uncertainties. The orange curve shows the PCA continuum absorbed by a proximate DLA in the foreground of the quasar (with $N_{\rm HI}=10^{20.5}$ cm$^{-2}$) centered on the vertical dotted line. The orange shaded regions show the propagated continuum uncertainty assuming a fixed DLA column density. The inset panel shows a segment of Ly$\alpha$ forest transmission (at the original spectral binning) at distances of 4--12 proper Mpc from the quasar assuming the PCA+DLA curve as the continuum level.}
\label{fig:ghosts_are_real}
\end{figure}

In Figure~\ref{fig:ghosts_are_real}, we show a zoom-in of the PSO J056--16 spectrum close to rest-frame Ly$\alpha$. The coincident redshift of a strong low-ionization metal absorber (vertical dotted line; to be discussed in Farina et al., in prep.) and a completely opaque trough in the immediate vicinity of the quasar (where there should otherwise be a strong proximity effect, e.g. \citealt{Davies19b}) both support the hypothesis that there is a foreground DLA. Beyond the DLA lie conspicuous transmission spikes in the Ly$\alpha$ forest. Measuring the Ly$\alpha$ forest transmission, however, requires an estimate of the unabsorbed quasar continuum. The broad Ly$\alpha$ emission line of PSO J056--16 is very strong, so the standard method of estimating the unabsorbed quasar continuum via extrapolating a power-law from the red side of the spectrum (e.g. \citealt{Fan06,Bosman18}), shown by the dashed blue curve in Figure~\ref{fig:ghosts_are_real}, is likely to substantially underestimate the true continuum level in the vicinity of the quasar.

To better estimate the unabsorbed quasar spectrum, we employed the principal component analysis (PCA) continuum reconstruction algorithm from \citet{Davies18a} to obtain the solid blue curve in Figure~\ref{fig:ghosts_are_real}. In short, we fit the PCA coefficients required to reproduce the spectrum red side of the region of interest (at rest wavelengths $1280 < \lambda_{\rm rest} < 2850$\,\AA), and then transformed the values of the red-side PCA coefficients to a prediction for the PCA coefficients of the blue side of the spectrum ($1175 < \lambda_{\rm rest} < 1280$\,\AA). The corresponding blue shaded regions show 1 and $2\sigma$ uncertainties determined from applying the same reconstruction process to similar SDSS/BOSS quasars (see \citealt{Davies18a} for details of this custom error estimation procedure and how ``similar" quasars are defined). The quasar spectrum close to rest-frame Ly$\alpha$ lies conspicuously below the continuum prediction, further suggesting the presence of a DLA, although the two are consistent at the $\sim2$--$3\sigma$ level taking into account the strong covariance in the continuum uncertainty. The orange curve shows the PCA continuum multiplied by a DLA at the redshift of the metal absorption with $N_{\rm HI}=10^{20.5}$ cm$^{-2}$ and the shaded regions show the propagated continuum uncertainty. Note that the DLA column density was chosen by eye to roughly match the recovery of the transmission on the red side of its opaque trough, and does not represent a quantitative fit. This PCA+DLA model then provides our estimate of the effective continuum level to measure the Ly$\alpha$ forest transmission.

The inset panel of Figure~\ref{fig:ghosts_are_real} shows the resulting Ly$\alpha$ forest transmission blueward of the saturated DLA trough, highlighting a range of foreground Ly$\alpha$ redshifts that cover distances of 4--12 proper Mpc from the quasar. Strong Ly$\alpha$ forest transmission spikes are apparent out to $\sim10$ proper Mpc, and the mean Ly$\alpha$ forest transmission from $z=5.8$--$5.9$ is $\langle F\rangle\approx0.13$, corresponding to an effective optical depth $\tau_{\rm eff}\approx2$. This the most Ly$\alpha$-transmissive large-scale region (outside of quasar proximity zones) ever observed at this redshift \citep{Eilers18,Bosman18}, and the $\sim5$--$10$ proper Mpc distance of the transmission from the quasar is more suggestive of the ghost proximity effect than the effect of halo bias\footnote{A careful reader will note that this level of transmission is even \emph{higher} than predicted for the ghost proximity effect in \S~\ref{sec:seminum_ghost} (Figure~\ref{fig:seminum_ghost_tra}). This may imply that the mean free path in the surrounding IGM is more sensitive to the quasar radiation than assumed in our ionizing background simulations (e.g. \citealt{McQuinn11}), the translation between photoionization rate and Ly$\alpha$ forest transmission that we assume may not be quite right, or it may represent a chance superposition of the ghost proximity effect with a pre-existing upward fluctuation.}.

\section{Conclusion} \label{sec:conc}

In this work, we explored the possibility of measuring the biased ionizing background around quasar-hosting dark matter halos in the shadow of proximate absorption systems, e.g. proximate DLAs. In principle, such a measurement jointly constrains the nature of the sources producing ionizing photons (e.g. \citetalias{Kakiichi18}) in addition to the identity of the quasar host halos themselves. Using a Gpc$^3$ volume semi-numerical simulation of $z=6$ ionizing background fluctuations, we showed that while the biased ionizing background around massive halos can be quite strong within a few proper Mpc, the observable signal in the Ly$\alpha$ forest is likely to be highly stochastic from sightline to sightline and, contrary to \citetalias{Kakiichi18}, we find that it should be strongly suppressed by the associated large-scale IGM overdensity. Further analysis with more realistic treatments (i.e. hydrodynamical simulations) of the IGM density and velocity structure within several proper Mpc of massive dark matter halos will be required to conclusively answer this question. 

We also showed that, despite its ionizing photons being extinguished along the line of sight, the quasar itself could still elevate the ionizing background along the line of sight via a ``ghost" of the classical proximity effect. While the absorber blocks ionizing photons along the line of sight, it likely only covers a very small solid angle of the quasar's total emission, and thus a very large volume surrounding the line of sight will still experience the intense ionizing radiation from the quasar. The mean free path of ionizing photons in the IGM surrounding the line of sight should then be relatively long compared to the typical IGM, so more ionizing photons will reach the line of sight, and thus increase the strength of the ionizing background. We denote this effect the ``ghost" proximity effect. The ghost proximity effect is the inevitable result of models in which the mean free path of ionizing photons is regulated by the strength of the ionizing background \citep{McQuinn11,Crociani11,DF16,D'Aloisio18}. From analytic arguments and realizations of quasar radiation in fluctuating ionizing background models, we showed that the strength of the ghost proximity effect would likely exceed the bias due to a massive host halo, and could in principle extend to distances beyond 10 proper Mpc along the line of sight. 

Observations of the Ly$\alpha$ forest in the immediate foreground of quasars with optically thick proximate absorption systems are required to investigate which, if any, of these effects are present. We showed a preliminary analysis of the $z\sim6$ quasar PSO J056--16 which has a proximate DLA, and which appears to show extraordinarily high Ly$\alpha$ forest transmission at distances of 5--10 proper Mpc in the shadow of the DLA. If the excess Ly$\alpha$ forest transmission is indeed due to the ghost proximity effect, then it should allow for strong constraints on the minimum lifetime and opening angle of emission from this quasar, but we leave a careful quantitative analysis for future work. In addition, the existence of the ghost proximity effect would imply that the mean free path is indeed biased high in the IGM within several proper Mpc of the quasar, motivating a re-analysis of the \citet{Worseck14} mean free path measurements from stacked quasar spectra at $z\sim5$ and their implications for the ionizing photon budget \citep{BB13,D'Aloisio18}, and potentially constraining the dependence of the mean free path on the strength of the UV background (as employed in, e.g., \citealt{DF16}).

Ongoing spectroscopic investigations of $z\ga6$ quasars are beginning to populate a statistical sample of proximate DLAs during the first billion years of cosmic time (\citealt{D'Odorico18,Banados19}; Farina et al., in prep.). Systematic study of the Ly$\alpha$ forest transmission beyond the saturated DLA troughs compared to average regions of the IGM will allow for quantitative constraints on the effects considered here, albeit likely with substantial degeneracies. While proximate DLAs are relatively easy to identify due to their extended damping wings (e.g. \citealt{D'Odorico18,Banados19}), the discovery of proximate super-Lyman limit systems (with $\tau_{\rm HI}\gtrsim10$, $N_{\rm HI}\gtrsim 10^{18.2}$ cm$^{-2}$) would saturate the minimum range of Ly$\alpha$ forest while still blocking the (direct) contribution of ionizing photons from the quasar.

\section*{Acknowledgements}

We thank Joe Hennawi, Vikram Khaire, and the rest of the ENIGMA group at UC Santa Barbara for valuable discussions and the computing resources used to generate the ionizing background models. We also thank Steve Furlanetto and Anson D'Aloisio for comments on a draft of this manuscript. FBD acknowledges support from the Space Telescope Science Institute, which is operated by AURA for NASA, through the grant HST-AR-15014.

\bibliographystyle{mnras}
 \newcommand{\noop}[1]{}

\end{document}